\documentclass[onecolumn,preprintnumbers,elsart]{revtex4}
\usepackage{eurosym}
\usepackage{makeidx}
\usepackage{amssymb}
\usepackage{amsmath}
\usepackage{mathrsfs}
\usepackage{graphicx}
\usepackage{dcolumn}
\usepackage{bm}
\usepackage[center]{subfigure}
\usepackage{color}

\begin{document}

\title{Spontaneous symmetry breaking of fundamental states, vortices, and
dipoles in two- and one-dimensional linearly coupled traps with cubic
self-attraction}
\author{Zhaopin Chen$^{1}$}
\author{Yongyao Li$^{2}$}
\author{Boris A. Malomed$^{1,3}$}
\author{Luca Salasnich$^{4.5}$}
\affiliation{$^{1}$Department of Physical Electronics, School of Electrical Engineering,
Faculty of Engineering, Tel Aviv University, Tel Aviv 69978, Israel\\
$^{2}$School of Physics and Optoelectronic Engineering, Foshan University,
Foshan 52800, China\\
$^{3}$Laboratory of Nonlinear-Optical Informatics, ITMO University,
St.Petersburg 197101, Russia\\
$^{4}$Dipartimento di Fisica e Astronomia ``Galileo Galilei" and CNISM,
Universit`a di Padova, via Marzolo 8, 35131 Padova, Italy\\
$^{5}$Istituto Nazionale di Ottica (INO) del Consiglio Nazionale delle
Ricerche (CNR), Sezione di Sesto Fiorentino, via Nello Carrara, 1 -- 50019
Sesto Fiorentino, Italy }

\begin{abstract}
We introduce two- and one-dimensional (2D and 1D) systems of two
linearly-coupled Gross-Pitaevskii equations (GPEs) with the cubic
self-attraction and harmonic-oscillator (HO) trapping potential in each GPE.
The system models a Bose-Einstein condensate with a negative scattering
length, loaded in a double-pancake trap, combined with the in-plane HO
potential. In addition to that, the 1D version applies to the light
transmission in a dual-core waveguide with the Kerr nonlinearity and in-core
confinement represented by the HO potential. The subject of the analysis is
spontaneous symmetry breaking in 2D and 1D ground-state (GS, alias
fundamental) modes, as well as in 2D vortices and 1D dipole modes (the
latter ones do not exist without the HO potential). By means of the
variational approximation and numerical analysis, it is found that both the
2D and 1D systems give rise to a symmetry-breaking bifurcation (SBB) of the
supercrtical type. Stability of symmetric states and asymmetric ones,
produced by the SBB, is analyzed through the computation of eigenvalues for
perturbation modes, and verified by direct simulations. The asymmetric GSs
are always stable, while the stability region for vortices shrinks and
eventually disappears with the increase of the linear-coupling constant, $%
\kappa $. The SBB in the 2D system does not occur if $\kappa $ is too large
(at $\kappa >\kappa _{\max }$); in that case, the two-component system
behaves, essentially, as its single-component counterpart. In the 1D system,
both asymmetric and symmetric dipole modes feature an additional oscillatory
instability, unrelated to the symmetry breaking. This instability occurs in
several regions, which expand with the increase of $\kappa $.
\end{abstract}

\maketitle

\section{Introduction}

A basic principle of the guided wave propagation in linear media is that the
ground state (GS) in such systems exactly follows the symmetry of the
guiding potential, while the first excited state features the opposite
parity. In particular, the GS of a quantum particle trapped in a double-well
potential is always symmetric, with respect to the potential structure,
while the wave function of the first excited state is antisymmetric \cite{LL}%
. Beyond the framework of the linear propagation, Bose-Einstein condensates
(BECs) are modeled by the Gross-Pitaevskii equation (GPE), which contains
the cubic term representing repulsive or attractive interactions between
atoms \cite{BEC}. Similar nonlinear Schr\"{o}dinger equations (NLSEs),
usually with the cubic self-focusing (Kerr) nonlinearity, are well known as
models of the guided light transmission in optics \cite{NLS}.

It is well known too that, if the GPE or NLSE contains a symmetric trapping
potential, the GS wave function follows its symmetry only if the strength of
the self-attractive nonlinearity does not exceed a certain critical level.
Above it, effects of the spontaneous symmetry breaking kick in,
destabilizing the symmetric state and replacing it, as the GS, by an
asymmetric wave function. These effects were originally predicted in early
works \cite{early}, and have then drawn much attention, due to their obvious
physical interest. In particular, they have been studied in detail in
one-dimensional (1D) dual-core systems, which represent nonlinear twin-core
optical waveguides \cite{bif1D}, modeled by a pair of linearly coupled
NLSEs. A similar system of 1D linearly coupled GPEs may be realized as the
model of a pair of parallel cigar-shaped traps, filled by self-attractive
BEC and linearly coupled by tunneling of atoms \cite{HS}. The latter model
is a natural extension of the single GPE with a double-well potential \cite%
{double-well-BEC}. In these systems, a basic problem is the spontaneous
symmetry breaking in two-component solitons (nonlinear confined modes),
through the phase transitions of the first or second kind, alias sub- and
supercritical bifurcations, respectively \cite{bif}. Experimentally,
spontaneous symmetry breaking has been demonstrated, in particular, in
lasing realized in various nonlinear-optical settings \cite{lasers}, in BEC
loaded in a double-well potential trap \cite{Markus}, and in optical
metamaterials \cite{Kivshar}.

Many theoretical and experimental results on the topic of the spontaneous
symmetry breaking in various nonlinear systems, chiefly in the 1D geometry,
have been collected in a recently published volume \cite{book} (see also a
review in Ref. \cite{review}). Fewer theoretical results have been reported
about symmetry-breaking bifurcations (SBBs) in 2D systems. In the absence of
trapping potentials, the SBB of 2D two-component solitons was studied in the
model of the spatiotemporal light propagation in a dual-core waveguide with
the intrinsic cubic-quintic nonlinearity \cite{Nir} (see also Ref. \cite%
{Gena}), which was adopted to prevent destruction of the solitons by the
collapse \cite{CQ}. The analysis was performed for fundamental solitons,
with zero vorticity ($S=0$), as well for the \textit{spatiotemporal vortex
solitons}, with $S=1$ (cf. Ref. \cite{X}, as concerns the latter concept).
The SBB\ for 2D\ fundamental and vortex solitons in a two-layer BEC,
supported by the lattice potential acting in both layers, was studied in
Ref. \cite{Arik}.

In this work, we aim to propose a sufficiently fundamental setting for the
study of the symmetry-breaking phenomenology in the 2D geometry: a system of
two linearly-coupled GPEs with the cubic self-attraction and an isotropic
harmonic-oscillator (HO) potential. It directly applies to the BEC with a
negative scattering, loaded in a combination of a planar double-pancake trap
and in-plane HO potential, the linear coupling being provided by tunneling
of atoms between the parallel \textquotedblleft pancakes". We also consider
a 1D version of this system, which additionally applies to the
spatial-domain light propagation in dual-core planar optical waveguides.
This setting was largely unexplored, except for Ref. \cite{sala-boris}.
However, the model considered in that work was essentially different, as it
was based on 2D mean-field equations for \emph{dense condensates}, which
include the GPE nonlinearly coupled to an additional equation for the
transverse width of the quasi-2D layer, i.e., the 2D version of the \textit{%
nonpolynomial Schr\"{o}dinger equation} (NPSE) \cite{Luca}. As a result, the
collapse area in the parameter space of the coupled NPSEs, found in Ref.
\cite{Luca}, is much larger than in the system of the coupled GPEs
considered here. Another essential difference is that the instability of
vortices against splitting was only mentioned, but not studied in Ref. \cite%
{Luca}, while the present paper studies it in detail, see Fig. \ref{SyAsyRegion}
below.

The model is introduced in Section II. It is followed by the consideration
of the symmetry-breaking of 2D confined states, with $S=0$ and $S=1$, in
Sections II and III, by means of the variational approximation (VA) and
numerical methods, respectively. The respective SBB is identified as a
supercritical one, and stability regions for symmetric and asymmetric 2D
fundamental and vortex modes are found. In the 1D system, the SBB and
stability are addressed for the GS solutions and dipole modes, i.e., the
lowest excited states; unlike the GSs and vortices, the 1D dipoles do not
exist in the absence of the HO potential. The paper is concluded by Section
IV.

\section{The model}

\subsection{The two-dimensional setting}

We consider the system of two linearly coupled 2D GPEs for complex wave
functions $\phi \left( x,y,t\right) $ and $\psi \left( x,y,t\right) $ with
the cubic self-attractive nonlinearity and HO trapping potential, written in
the usual scaled form:

\begin{eqnarray}
i\phi _{t} &=&-\frac{1}{2}\left( \phi _{xx}+\phi _{yy}\right) -|\phi
|^{2}\phi +\frac{1}{2}\left( x^{2}+y^{2}\right) \phi -\kappa \psi ,
\label{phi} \\
i\psi _{t} &=&-\frac{1}{2}\left( \psi _{xx}+\psi _{yy}\right) -|\psi
|^{2}\psi +\frac{1}{2}\left( x^{2}+y^{2}\right) \psi -\kappa \phi .
\label{psi}
\end{eqnarray}%
The coefficients in front of the Laplacian, cubic term, and HO potential are
set equal to $1$ by means of rescaling, the single irreducible parameter
being the linear-coupling coefficient, $\kappa >0$ (naturally, we assume
that the scattering lengths, i.e., strengths of the nonlinear terms, are
equal in both components). Equations (\ref{phi}) and (\ref{psi}) can be
easily derived from the underlying 3D GPE by imposing a strong confinement
along the third direction ($z$ axis), corresponding to the double-pancake
configuration, and a factorized Gaussian wave function in the $z$ direction
around each local maximum, with the width equal to the characteristic
harmonic length \cite{sala-boris}.

It is relevant to relate scaled units in which Eqs. (\ref{phi}) and (\ref%
{psi}) are written to their physical counterparts. Assuming the gas of $^{7}$%
Li atoms, with scattering length $\sim -0.1$ nm, which typically corresponds
to the attractive interactions \cite{Hulet}, the transverse confinement
provided by the harmonic-oscillator potential with strength $\omega _{\perp
}\sim 10$ KHz, and in-plane potential [in Eqs. (\ref{phi}) and (\ref{psi})]
with much smaller strength, $\Omega \sim 100$ Hz, the corresponding
transverse and in-plane confining radii being, respectively, $r_{\perp }\sim
1$ $\mathrm{\mu }$m and $10$ $\mathrm{\mu }$m, the scaled length and time
units in Eqs. (\ref{phi}) and (\ref{psi}) translate into $\sim 10$ $\mathrm{%
\mu }$m and $100$ ms, respectively. Thus, typical radii of the fundamental
and vortex modes presented below, which are $\sim 3$ in the scaled units,
correspond $\sim 30$ $\mathrm{\mu }$m in physical units. Further, the actual
number of atoms in the condensate, $\mathcal{N}$, is related to the scaled
norm of the wave function [see Eq. (\ref{Ncr}) below] by formula\
\begin{equation}
\mathcal{N}\sim 10^{3}N,  \label{NN}
\end{equation}%
hence the SBB, which is shown below to take place at $N\gtrsim 3$, implies
that the condensate must contain, at least, $\sim 3000$ atoms. In the 1D
case considered below, estimates for the physical time and length units are
essentially the same, while the relation between the actual number of atoms
in the quasi-1D condensate and its scaled norm is $\mathcal{N}\sim 10^{2}N$,
instead of Eq. (\ref{NN}).

Stationary solutions to Eqs. (\ref{phi}) and (\ref{psi}) with real chemical
potential $\mu $ are looked for as%
\begin{equation}
\left\{ \phi ,\psi \right\} =e^{-i\mu t+iS\theta }\left\{ \Phi _{S}(r),\Psi
_{S}(r)\right\} ,  \label{PhiPsi}
\end{equation}%
where $\left( r,\theta \right) $ are the polar coordinates, $S=0,1,2,...$ is
the vorticity \cite{old}, and real functions $\Phi _{S}$ and $\Psi _{S}$ are
determined as solutions of coupled ordinary differential equations, with the
prime standing for $d/dr$:%
\begin{eqnarray}
\mu \Phi _{S} &=&-\frac{1}{2}\left( \Phi _{S}^{\prime \prime }+\frac{1}{r}%
\Phi _{S}^{\prime }-\frac{S^{2}}{r^{2}}\Phi _{S}\right) -\Phi _{S}^{3}+\frac{%
1}{2}r^{2}\Phi _{S}-\kappa \Psi _{S},  \label{Phi} \\
\mu \Psi _{S} &=&-\frac{1}{2}\left( \Psi _{S}^{\prime \prime }+\frac{1}{r}%
\Psi _{S}^{\prime }-\frac{S^{2}}{r^{2}}\Psi _{S}\right) -\Psi _{S}^{3}+\frac{%
1}{2}r^{2}\Psi _{S}-\kappa \Phi _{S}.  \label{Psi}
\end{eqnarray}%
Symmetric solutions, with $\phi =\psi $, are governed by the single equation,%
\begin{equation}
i\phi _{t}=-\frac{1}{2}\left( \phi _{xx}+\phi _{yy}\right) -|\phi |^{2}\phi +%
\frac{1}{2}\left( x^{2}+y^{2}\right) \phi -\kappa \phi .  \label{single}
\end{equation}%
On the other hand, in the limit of $\mu \rightarrow -\infty $ the small
component of an extremely asymmetric solution is given by the relation
following from Eq. (\ref{Psi}),%
\begin{equation}
\Psi _{S}\approx -\left( \kappa /\mu \right) \Phi _{S},  \label{small}
\end{equation}%
with solution $\Phi _{S}$ given by Eq. (\ref{Phi}) in which term $\kappa
\Psi _{S}$ is omitted.

A crucially important issue is stability of solutions to Eqs. (\ref{phi})
and (\ref{psi}). In particular, antisymmetric states, with $\phi =-\psi $,
are unstable, as $\kappa >0$ implies that they have a positive, rather than
negative, coupling energy. Further, known facts are that all the solutions
of the single equation (\ref{single}) with $S=0$ are stable, and a part of
vortices with $S=1$ are stable too, while all vortices with $S\geq 2$ are
unstable \cite{Dum,HSaito}. Therefore, in what follows below we consider
only two cases, $S=0$ (fundamental modes) and $S=1$ (unitary vortices).

The basic objective of the analysis is to find an SBB which destabilizes the
symmetric states and replaces them by nontrivial stable asymmetric ones, if
the total norm of the given state, $N$, exceeds a certain critical value:%
\begin{equation}
N=N_{\phi }+N_{\psi }\equiv 2\pi \int_{0}^{\infty }\Phi _{S}^{2}(r)rdr+2\pi
\int_{0}^{\infty }\Psi _{S}^{2}(r)rdr>N_{\mathrm{cr}}^{(S)}(\kappa ).
\label{Ncr}
\end{equation}%
In fact, it is necessary to find $N_{\mathrm{cr}}^{(S)}$ for $S=0$ and $1$
as functions of $\kappa $, and investigate the character of the SBB, which
is determined by dependence of the respective asymmetry parameter,%
\begin{equation}
\theta \equiv \frac{\left\vert N_{\psi }-N_{\phi }\right\vert }{N},
\label{theta}
\end{equation}%
on $N$, for given $S$ and $\kappa $.

To analyze stability of the stationary states, we search for perturbed
solutions to Eqs.(\ref{phi}) and (\ref{psi}) as
\begin{eqnarray}
\phi &=&\left[ \Phi _{S}(r)+u_{1}(r)e^{\lambda t+iL\theta }+u_{2}^{\ast
}(r)e^{\lambda ^{\ast }t-iL\theta }\right] e^{iS\theta -i\mu t},
\label{PerSolphi} \\
\psi &=&\left[ \Psi _{S}(r)+v_{1}(r)e^{\lambda t+iL\theta }+v_{2}^{\ast
}(r)e^{\lambda ^{\ast }t-iL\theta }\right] e^{iS\theta -i\mu t},
\label{PerSolpsi}
\end{eqnarray}%
where $u_{1,2}(x,z)$ and $v_{1,2}(x,z)$ are perturbation eigenmodes with
integer azimuthal index $L$, and $\lambda $ is the corresponding (generally,
complex) instability growth rate. Linearization around the stationary
solutions leads to the Bogoliubov - de Gennes (BdG) equations:
\begin{eqnarray}
-\frac{1}{2}(u_{1}^{\prime \prime }+\frac{1}{r}u_{1}^{\prime }-\frac{%
(S+L)^{2}}{r^{2}}u_{1})-\Phi ^{2}(2u_{1}+u_{2})+\frac{1}{2}r^{2}u_{1}-\kappa
v_{1}-\mu u_{1} &=&i\lambda u_{1},  \notag \\
-\frac{1}{2}(u_{2}^{\prime \prime }+\frac{1}{r}u_{2}^{\prime }-\frac{%
(S-L)^{2}}{r^{2}}u_{2})-\Phi ^{2}(2u_{2}+u_{1})+\frac{1}{2}r^{2}u_{2}-\kappa
v_{2}-\mu u_{2} &=&-i\lambda u_{2},  \notag \\
-\frac{1}{2}(v_{1}^{\prime \prime }+\frac{1}{r}v_{1}^{\prime }-\frac{%
(S+L)^{2}}{r^{2}}v_{1})-\Psi ^{2}(2v_{1}+v_{2})+\frac{1}{2}r^{2}v_{1}-\kappa
u_{1}-\mu v_{1} &=&i\lambda v_{1},  \notag \\
-\frac{1}{2}(v_{2}^{\prime \prime }+\frac{1}{r}v_{2}^{\prime }-\frac{%
(S-L)^{2}}{r^{2}}v_{2})-\Psi ^{2}(2v_{2}+v_{1})+\frac{1}{2}r^{2}v_{2}-\kappa
u_{2}-\mu v_{2} &=&-i\lambda v_{2},  \label{eigfunct}
\end{eqnarray}
which were solved numerically, with boundary condition demanding $u(r)$ and $%
v(r)$ to decay as $r^{|S\pm L|}$ at $r\rightarrow 0$, and exponentially at $%
r\rightarrow \infty $. The instability is predicted by the existence of
(pairs of) eigenvalues with $\mathrm{Re}(\lambda )\neq 0$. In particular,
all unstable eigenvalues which account for the symmetry-breaking instability
in the 2D and 1D systems considered below are (as usual) purely real, i.e.,
the corresponding instability is not oscillatory. Complex eigenvalues are
found in the case of unstable 2D vortices (shown below in Figs. \ref%
{SrcInstb} and \ref{AsyInstb}), and specific instability (which is unrelated
to the symmetry breaking) of 1D dipole modes, see Figs. \ref{1dNmu}(d) and %
\ref{AsyN1dOdd} below.

\subsection{Reduction to one dimension}

The 1D version of Eqs. (\ref{phi})-(\ref{psi}) is mathematically obtained by
dropping all terms containing $y$:

\begin{eqnarray}
i\phi _{t} &=&-\frac{1}{2}\phi _{xx}-|\phi |^{2}\phi +\frac{1}{2}x^{2}\phi
-\kappa \psi ,  \label{phi1D} \\
i\psi _{t} &=&-\frac{1}{2}\psi _{xx}-|\psi |^{2}\psi +\frac{1}{2}x^{2}\psi
-\kappa \phi .  \label{psi1D}
\end{eqnarray}
In terms of BEC, Eqs. (\ref{phi1D}) and (\ref{psi1D}) can be derived from
Eqs. (\ref{phi}) and (\ref{psi}), assuming a strong confinement along the $y$
direction and the corresponding factorization of the wave function, with the
Gaussian accounting for its structure in the $y$-direction. In addition to
the BEC loaded in the double trap, this system applies, as mentioned above,
to optics as well: with $t$ replaced by propagation distance $z$, it models
the propagation of light in a dual-core planar waveguide with the intrinsic
Kerr nonlinearity and trapping potential representing a guiding channel. In
addition to the study of the symmetry breaking of fundamental (spatially
even) modes, in the 1D case it is also relevant to address the same effect
featured by dipole (spatially odd) modes, i.e., the lowest excited state, in
terms of quantum mechanics. Note that, unlike the soliton-like fundamental
states, which exist in the free 1D space, dipole modes do not exist in the
absence of the trapping potential.

Stationary solutions to Eqs. (\ref{phi}) and (\ref{psi}) with real chemical
potential $\mu $ are looked for as%
\begin{equation}
\left\{ \phi ,\psi \right\} =e^{-i\mu t}\left\{ \Phi (x),\Psi (x)\right\} ,
\label{mu1D}
\end{equation}%
where real functions $\Phi $ and $\Psi $ are determined as solutions of
coupled ordinary differential equations, with the prime standing for $d/dx$:%
\begin{eqnarray}
\mu \Phi &=&-\frac{1}{2}\Phi ^{\prime \prime }-\Phi ^{3}+\frac{1}{2}%
x^{2}\Phi -\kappa \Psi ,  \label{Phi1D} \\
\mu \Psi &=&-\frac{1}{2}\Psi ^{\prime \prime }-\Psi ^{3}+\frac{1}{2}%
x^{2}\Psi -\kappa \Phi ,  \label{Psi1D}
\end{eqnarray}
cf. Eqs. (\ref{PhiPsi})-(\ref{Psi}). Symmetric solutions, with $\phi =\psi $%
, obey the single equation, which is the 1D version of Eq. (\ref{single}):%
\begin{equation}
i\phi _{t}=-\frac{1}{2}\phi _{xx}-|\phi |^{2}\phi +\frac{1}{2}x^{2}\phi
-\kappa \phi .  \label{single1D}
\end{equation}%
Different solutions are characterized by their norm (in the application to
optics, it is the total power of the two-component optical beam), which is
defined by the 1D counterpart of Eq. (\ref{Ncr}): $N=N_{\phi }+N_{\psi
}\equiv \int_{-\infty }^{+\infty }\Phi ^{2}(x)dx+\int_{-\infty }^{+\infty
}\Psi ^{2}(x)dx$. At $N$ exceeding the respective critical value, $N_{%
\mathrm{cr}}^{(\mathrm{1D})}(\kappa )$, the asymmetry is defined as per the
above equation (\ref{theta}). The stability of 1D modes was investigated by
means of the respectively simplified system of BdG equations (\ref{eigfunct}%
).

\section{The analytical approach: the variational approximation}

\subsection{The two-dimensional system}

A natural analytical approach to the solution of Eqs. (\ref{Phi}) and (\ref%
{Psi}) is based on the VA. Here, it is presented for $S=0$; for $S=1$ it can
be elaborated too, in a more cumbersome form. The variational ansatz is
adopted as the GS wave function of the 2D HO potential corresponding to Eqs.
(\ref{Phi}) and (\ref{Psi}) [here, $\Phi _{S}(r)$ and $\Psi _{S}(r)$ are
replaced by $\Phi (r)$ and $\Psi (r)$]:%
\begin{equation}
\left\{ \Phi (r),\Psi (r)\right\} =\left\{ A,B\right\} \exp \left( -\frac{1}{%
2}r^{2}\right) ,  \label{ans}
\end{equation}%
with unknown amplitudes (variational parameters) $A$ and $B$, the norm of
this ansatz being
\begin{equation}
N^{\mathrm{(2D)}}\equiv N_{\phi }^{\mathrm{(2D)}}+N_{\psi }^{\mathrm{(2D)}%
}=\pi A^{2}+\pi B^{2}.  \label{N2D}
\end{equation}%
This ansatz implies that the nonlinearity is not too strong, as, otherwise,
the nonlinear self-focusing essentially alters the shape of the 2D trapped
state, switching it from the GS of the HO potential towards a \textit{Townes
soliton }\cite{Berge}, for which the VA is built differently, using both the
amplitude and width of the mode as variational parameters \cite{Anderson}
(actually, this version of the VA was developed only for the
single-component GPE in the absence of the trapping potential). In
particular, strong self-focusing makes the width of the localized modes
smaller than the HO width, which is implied in ansatz (\ref{ans}); the same
pertains to the 1D \textit{ans\"{a}tze}, which are adopted below in Eqs. (%
\ref{ansatz1D}) and (\ref{ansatz-odd}). We do not aim to develop such an
improved version of the VA in the present work, as the resulting algebra is
rather cumbersome.

The VA is based on the Lagrangian corresponding to Eqs. (\ref{Phi}) and (\ref%
{Psi}):
\begin{equation}
L=\int_{0}^{\infty }rdr\left[ \frac{1}{4}\left( \left( \Phi ^{\prime
}\right) ^{2}+\left( \Psi ^{\prime }\right) ^{2}\right) -\frac{1}{4}\left(
\Phi ^{4}+\Psi ^{4}\right) +\frac{1}{4}r^{2}\left( \Phi ^{2}+\Psi
^{2}\right) -\kappa \Phi \Psi -\frac{\mu }{2}\left( \Phi ^{2}+\Psi
^{2}\right) \right] .  \label{L}
\end{equation}%
The substitution of ansatz (\ref{ans}) in Lagrangian (\ref{L}) yields the
following effective Lagrangian, as a function of $A$ and $B$:%
\begin{equation}
L_{\mathrm{eff}}=\frac{1}{4}(1-\mu )(A^{2}+B^{2})-\frac{1}{16}\left(
A^{4}+B^{4}\right) -\frac{\kappa }{2}AB.  \label{Leff}
\end{equation}%
The corresponding Euler-Lagrange equations, $\partial L_{\mathrm{eff}%
}/\partial A=\partial L_{\mathrm{eff}}/\partial B=0$, amount to a system of
coupled cubic equations:%
\begin{eqnarray}
\left( 1-\mu \right) A-\kappa B-\frac{1}{2}A^{3} &=&0,  \label{A} \\
\left( 1-\mu \right) B-\kappa A-\frac{1}{2}B^{3} &=&0.  \label{B}
\end{eqnarray}%
The symmetric solution of these equations is obvious:%
\begin{equation}
A^{2}=B^{2}=2\left( 1-\kappa -\mu \right) ,  \label{symm}
\end{equation}%
the corresponding norm (\ref{N2D}) being%
\begin{equation}
N_{\mathrm{symm}}^{\mathrm{(2D)}}=4\pi \left( 1-\kappa -\mu \right) .
\label{VAN2D}
\end{equation}%
The solution exists at $N^{\mathrm{(2D)}}\geq 0$, i.e.,
\begin{equation}
\mu \leq \mu _{\max }\equiv 1-\kappa  \label{mumax}
\end{equation}%
(note that $1-$ $\kappa $ may be both positive and negative).

The asymmetric solution of Eqs. (\ref{A}) and (\ref{B}) can also be found in
an exact form:%
\begin{equation}
A^{2}=1-\mu +\sqrt{\left( 1-\mu \right) ^{2}-4\kappa ^{2}},~B^{2}=1-\mu -%
\sqrt{\left( 1-\mu \right) ^{2}-4\kappa ^{2}},  \label{asymm}
\end{equation}%
with the total norm%
\begin{equation}
N_{\mathrm{asymm}}^{\mathrm{(2D)}}=2\pi \left( 1-\mu \right) \text{.}
\label{VAN2Dasymm}
\end{equation}%
As it follows from Eqs. (\ref{mumax}) and (\ref{asymm}), only the symmetric
solution exists in interval
\begin{equation}
1-2\kappa <\mu \leq 1-\kappa .  \label{symmetric}
\end{equation}%
The asymmetric solution appears at
\begin{equation}
\mu =\mu _{\mathrm{cr}}^{(S=0)}\equiv 1-2\kappa ,~N=N_{\mathrm{cr}%
}^{(S=0)}\equiv 4\pi \kappa ,  \label{cr-2D}
\end{equation}%
and exists at $\mu <1-2\kappa $.

The asymmetric solution (\ref{asymm}) can be rewritten in terms of the
asymmetry parameter, defined by Eq. (\ref{theta}), and norm $N$ [see Eq. (%
\ref{N2D})]:%
\begin{equation}
\theta _{\mathrm{VA}}=\sqrt{1-\left( \frac{4\pi \kappa }{N}\right) ^{2}}
\label{thetaVA}
\end{equation}%
With the increase of $N$, the asymmetric solutions appears at $N=N_{\mathrm{%
cr}}^{(S=0)}$, and $\theta $ grows monotonously as a function of $N$ at $%
N>N_{\mathrm{cr}}^{(S=0)}$. Thus, VA predicts the SBB of the supercritical
type.

\subsection{The one-dimensional case}

\subsubsection{The ground state}

The VA ansatz for the 1D GS is adopted as the GS wave function for the 1D HO
potential corresponding to Eqs. (\ref{Phi1D}) and (\ref{Psi1D}):%
\begin{equation}
\left\{ \Phi (r),\Psi (r)\right\} =\left\{ A,B\right\} \exp \left( -\frac{1}{%
2}x^{2}\right) ,  \label{ansatz1D}
\end{equation}%
with variational parameters $A$ and $B$ and the respective norm,
\begin{equation}
N^{\mathrm{(1D)}}\equiv N_{\phi }^{\mathrm{(1D)}}+N_{\psi }^{\mathrm{(1D)}}=%
\sqrt{\pi }A^{2}+\sqrt{\pi }B^{2}.  \label{Norm1D}
\end{equation}%
The VA is based on the Lagrangian corresponding to Eqs. (\ref{Phi1D}) and (%
\ref{Psi1D}):%
\begin{equation}
L=\int_{-\infty }^{+\infty }dx\left[ \frac{1}{4}\left( \left( \Phi ^{\prime
}\right) ^{2}+\left( \Psi ^{\prime }\right) ^{2}\right) -\frac{1}{4}\left(
\Phi ^{4}+\Psi ^{4}\right) +\frac{1}{4}x^{2}\left( \Phi ^{2}+\Psi
^{2}\right) -\kappa \Phi \Psi -\frac{\mu }{2}\left( \Phi ^{2}+\Psi
^{2}\right) \right] ,  \label{L1D}
\end{equation}%
cf. Eq. (\ref{L}). The substitution of ansatz (\ref{ansatz1D}) in Lagrangian
(\ref{L1D}) yields%
\begin{equation}
\frac{1}{\sqrt{\pi }}L_{\mathrm{eff}}=\frac{1}{2}\left( \frac{1}{2}-\mu
\right) (A^{2}+B^{2})-\frac{1}{4\sqrt{2}}\left( A^{4}+B^{4}\right) -\kappa
AB.  \label{Leff1D}
\end{equation}%
The corresponding Euler-Lagrange equations, $\partial L_{\mathrm{eff}%
}/\partial A=\partial L_{\mathrm{eff}}/\partial B=0$, amount to a system of
coupled cubic equations [cf. Eqs. (\ref{A}) and (\ref{B}) in the 2D system]:%
\begin{eqnarray}
\left( \frac{1}{2}-\mu \right) A-\kappa B-\frac{1}{\sqrt{2}}A^{3} &=&0,
\label{A1D} \\
\left( \frac{1}{2}-\mu \right) B-\kappa A-\frac{1}{\sqrt{2}}B^{3} &=&0.
\label{B1D}
\end{eqnarray}%
The symmetric solution of these equations is%
\begin{equation}
A^{2}=B^{2}=\sqrt{2}\left( \frac{1}{2}-\mu -\kappa \right) ,  \label{symm1D}
\end{equation}%
the respective total norm (\ref{Norm1D}) being%
\begin{equation}
N_{\mathrm{symm}}^{\mathrm{(1D)}}=\sqrt{2\pi }\left( 1-2\mu -2\kappa \right)
.  \label{VAN1D}
\end{equation}%
It exists at
\begin{equation}
\mu \leq \mu _{\max }^{\mathrm{(1D)}}\equiv \frac{1}{2}-\kappa .
\label{mumax1D}
\end{equation}

The asymmetric solution of Eqs. (\ref{A}) and (\ref{B}) can be found in an
exact form too, cf. Eq. (\ref{asymm}):%
\begin{equation}
A^{2}=\frac{1-2\mu }{2\sqrt{2}}+\sqrt{\frac{\left( 1-2\mu \right) ^{2}}{8}%
-2\kappa ^{2}},~B^{2}=\frac{1-2\mu }{2\sqrt{2}}-\sqrt{\frac{\left( 1-2\mu
\right) ^{2}}{8}-2\kappa ^{2}},  \label{asymm1D}
\end{equation}%
with the total norm%
\begin{equation}
N_{\mathrm{asymm}}^{\mathrm{(1D)}}=\frac{1}{2}\sqrt{2\pi }\left( 1-2\mu
\right) .  \label{VAN1Dasymm}
\end{equation}%
As it follows from Eqs. (\ref{mumax}) and (\ref{asymm}), in interval
\begin{equation}
\frac{1}{2}-2\kappa <\mu \leq \frac{1}{2}-\kappa  \label{symmetric1D}
\end{equation}%
only the symmetric solution exists. The asymmetric one appears at
\begin{equation}
\mu =\mu _{\mathrm{cr}}^{\mathrm{(1D)}}=\frac{1}{2}-2\kappa ,~N=N_{\mathrm{cr%
}}^{\mathrm{(1D)}}=2\sqrt{2\pi }\kappa ,  \label{cr-1D}
\end{equation}%
and exists at $\mu \leq \frac{1}{2}-2\kappa $.

Asymmetric solution (\ref{asymm}) is characterized by the asymmetry
parameter, defined by Eq. (\ref{theta}), as a function of the total norm, $N$
[see Eq. (\ref{Norm1D})]:%
\begin{equation}
\theta _{\mathrm{VA}}^{\mathrm{(1D)}}=\sqrt{1-\frac{8\pi \kappa ^{2}}{N^{2}}}%
.  \label{thetaVA1D}
\end{equation}%
Thus, with the increase of $N$, the asymmetric solution appears at $N=N_{%
\mathrm{cr}}^{\mathrm{(1D)}}$, and the corresponding SBB\ is of the
supercritical type, as well as predicted above in the 2D system, on the
contrary to the well-known \textit{weakly subcritical} bifurcation for 1D
solitons in the free space \cite{bif1D}.

\subsubsection{The dipole mode}

The consideration of the SBB\ in the dipole mode, is especially interesting,
as such a mode, unlike the even GS soliton, does not exist in the absence of
the HO trapping potential. The corresponding VA ansatz is naturally adopted
as [cf. Eq. (\ref{ansatz1D})]%
\begin{equation}
\left\{ \Phi (r),\Psi (r)\right\} =\left\{ A,B\right\} ~x\exp \left( -\frac{1%
}{2}x^{2}\right) ,  \label{ansatz-odd}
\end{equation}%
its norms being [cf. Eq. (\ref{Norm1D})]%
\begin{equation}
N_{\phi }=\frac{1}{2}\sqrt{\pi }A^{2},~N_{\psi }=\frac{1}{2}\sqrt{\pi }B^{2}.
\label{Nodd1D}
\end{equation}

The calculation of Lagrangian (\ref{L1D}) with ansatz (\ref{ansatz-odd})
yields%
\begin{equation}
\frac{1}{\sqrt{\pi }}L_{\mathrm{eff}}^{\mathrm{(odd)}}=\frac{1}{4}\left(
\frac{3}{2}-\mu \right) (A^{2}+B^{2})-\frac{3}{64\sqrt{2}}\left(
A^{4}+B^{4}\right) -\frac{1}{2}\kappa AB,  \label{Leff-odd}
\end{equation}%
cf. Eq. (\ref{Leff1D}). Finally, the solution of the corresponding
Euler-Lagrange equation for $A$ and $B$ yields the following results, cf.
Eqs. (\ref{symm1D}) and (\ref{asymm1D}) for the GS. The symmetric solution is%
\begin{equation}
A_{\mathrm{odd}}^{2}=B_{\mathrm{odd}}^{2}=\frac{8\sqrt{2}}{3}\left( \frac{3}{%
2}-\mu -\kappa \right) ,  \label{symm-odd}
\end{equation}%
the respective total norm being
\begin{equation}
N_{\mathrm{odd}}^{\mathrm{(symm)}}\equiv N_{\phi }+N_{\psi }=\frac{8}{3}%
\sqrt{2\pi }\left( \frac{3}{2}-\mu -\kappa \right) ,  \label{Nodd1Dtotal}
\end{equation}%
as per Eq. (\ref{Nodd1D}). Accordingly, this state exists at
\begin{equation}
\mu \leq \mu _{\max }^{\mathrm{(1D,odd)}}\equiv \frac{3}{2}-\kappa .
\label{mumax-odd}
\end{equation}%
The asymmetric solution is%
\begin{eqnarray}
A_{\mathrm{odd}}^{2} &=&\frac{8}{3}\left[ \frac{3-2\mu }{2\sqrt{2}}+\sqrt{%
\frac{\left( 3-2\mu \right) ^{2}}{8}-2\kappa ^{2}}\right] ,  \notag \\
~B_{\mathrm{odd}}^{2} &=&\frac{8}{3}\left[ \frac{3-2\mu }{2\sqrt{2}}-\sqrt{%
\frac{\left( 3-2\mu \right) ^{2}}{8}-2\kappa ^{2}}\right] ,
\label{asymm-odd}
\end{eqnarray}%
with the total norm%
\begin{equation}
N_{\mathrm{odd}}^{\mathrm{(asymm)}}=\frac{2}{3}\sqrt{2\pi }\left( 3-2\mu
\right) ,  \label{Nodd1Dasymm}
\end{equation}%
cf. Eq. (\ref{Nodd1Dtotal}). As it follows from here, only the symmetric
dipole mode exists in interval%
\begin{equation}
\frac{3}{2}-2\kappa <\mu \leq \frac{3}{2}-\kappa ,  \label{symmetric-odd}
\end{equation}%
cf. Eq. (\ref{symmetric1D}). Its asymmetric counterpart appears at $\mu =%
\frac{3}{2}-2\kappa ,$ existing at $\mu \leq \frac{3}{2}-2\kappa $.

The asymmetric solution (\ref{asymm-odd}) can be rewritten in terms of the
asymmetry parameter defined by Eq. (\ref{theta}), and norm $N$, see Eq. (\ref%
{Nodd1D}):%
\begin{equation}
\theta _{\mathrm{VA}}^{\mathrm{(1D,odd)}}=\sqrt{1-\frac{128\pi \kappa ^{2}}{%
9N^{2}}}.  \label{thetaVA-odd}
\end{equation}%
Thus, with the increase of $N$, the asymmetric dipole mode appears at
\begin{equation}
N_{\mathrm{cr}}^{\mathrm{(1D,odd)}}=\frac{8\sqrt{2\pi }}{3}\kappa ,
\label{Ncr-odd}
\end{equation}%
and $\theta $ grows monotonously as a function of $N$ at $N>N_{\mathrm{cr}}^{%
\mathrm{(1D,odd)}}$, the corresponding SBB\ again being of the supercritical
type.

\section{Numerical results}

\subsection{The two-dimensional system}

\subsubsection{Symmetric and asymmetric ground-state (GS) modes with zero
vorticity}

GS solutions to Eqs. (\ref{Phi}) and (\ref{Psi}) with $S=0$ were produced by
means of the imaginary-time-integration method \cite{IT}, applied to
underlying equations (\ref{phi}) and (\ref{psi}). Then, the stability of
these solutions was identified through the computation of their eigenvalue
spectra as per Eq. (\ref{eigfunct}), and further verified by direct
simulations of the perturbed evolution. The eigenvalue spectra were explored
for perturbations with $L=0$, $L=1$ and $L=2$. For the GSs, the
perturbations with azimuthal index $L=0$ in Eqs. (\ref{PerSolphi}) and (\ref%
{PerSolpsi}) are found, quite naturally, to be the most dangerous ones,
while for the vortices with $S=1$, the stability boundaries are always
determined by the eigenmodes with $L=1$ or $L=2$, which is natural too \cite%
{Dum}. To test the stability in direct simulations, random noise at the $5\%$
amplitude level was locally added to initial conditions.

\begin{figure}[tbp]
\centering{\label{fig1a} \includegraphics[scale=0.23]{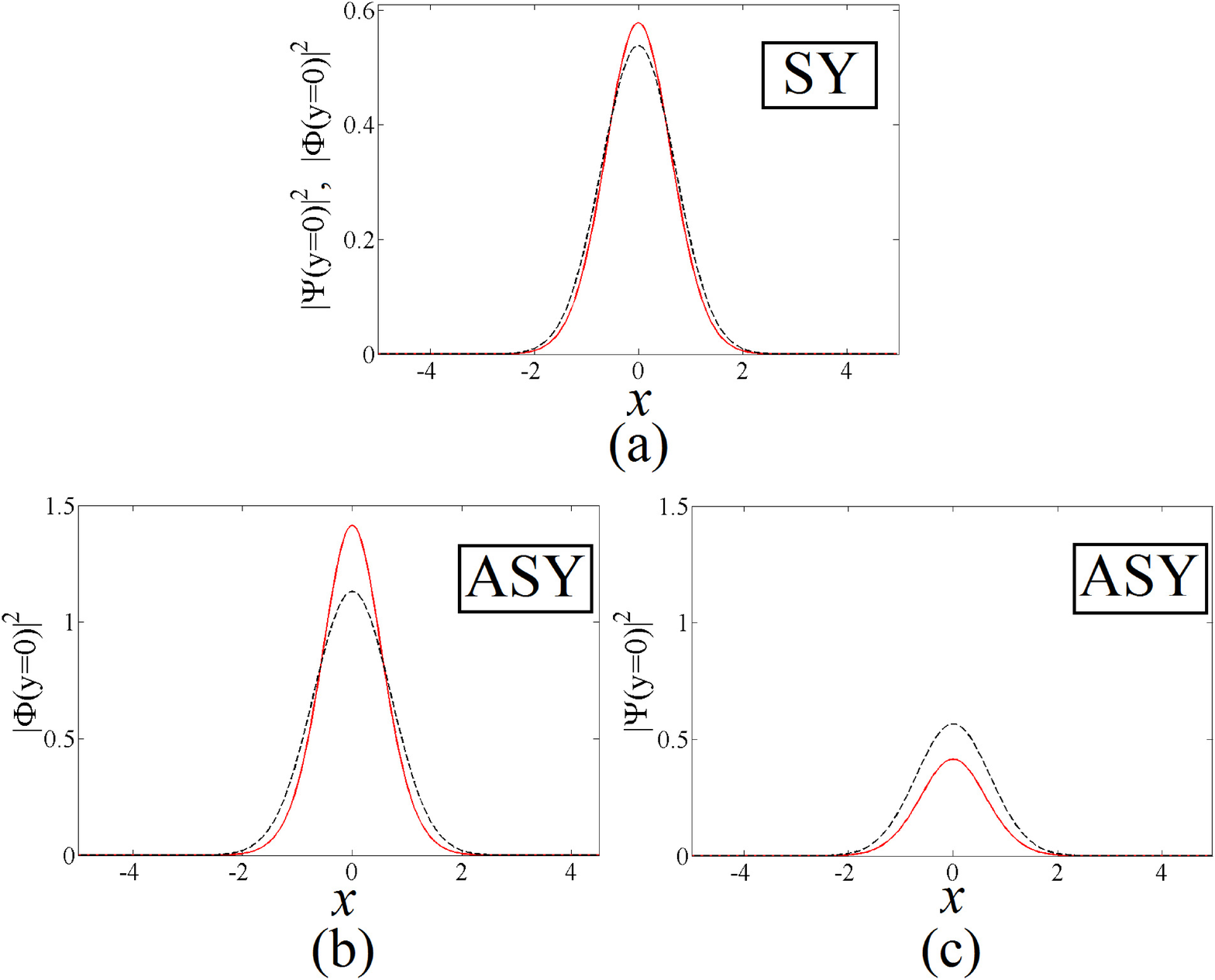}}
\caption{(Color online) (a) The cross section of the 2D profile of a stable
symmetric ground state with $(\protect\kappa ,N)=(0.4,1)$. (b,c) The same
for a stable asymmetric ground state with $(\protect\kappa ,N)=(0.4,4)$. Red
solid and black dashed curves designate numerical and VA results,
respectively. }
\label{syexample}
\end{figure}

First, in Fig. \ref{syexample} we display typical examples of the
numerically found stable symmetric and asymmetric 2D fundamental states
(being stable solutions, they represent the system's GS), along with their
counterparts predicted by the VA, which is based on Eqs. (\ref{ans}), (\ref%
{symm}), and (\ref{asymm}), respectively. The comparison of the numerical
and VA solutions is presented for identical values of the total norm. The
numerical solutions for the symmetric states, with $\Phi =\Psi $, coincide
with their counterparts previously produced by the single-component equation
(\ref{single}) with $\kappa =0$ \cite{Dum,HSaito} (of course, the stability
of the symmetric states may be different in the single- and two-component
systems).

\begin{figure}[tbp]
\centering{\label{fig2a} \includegraphics[scale=0.15]{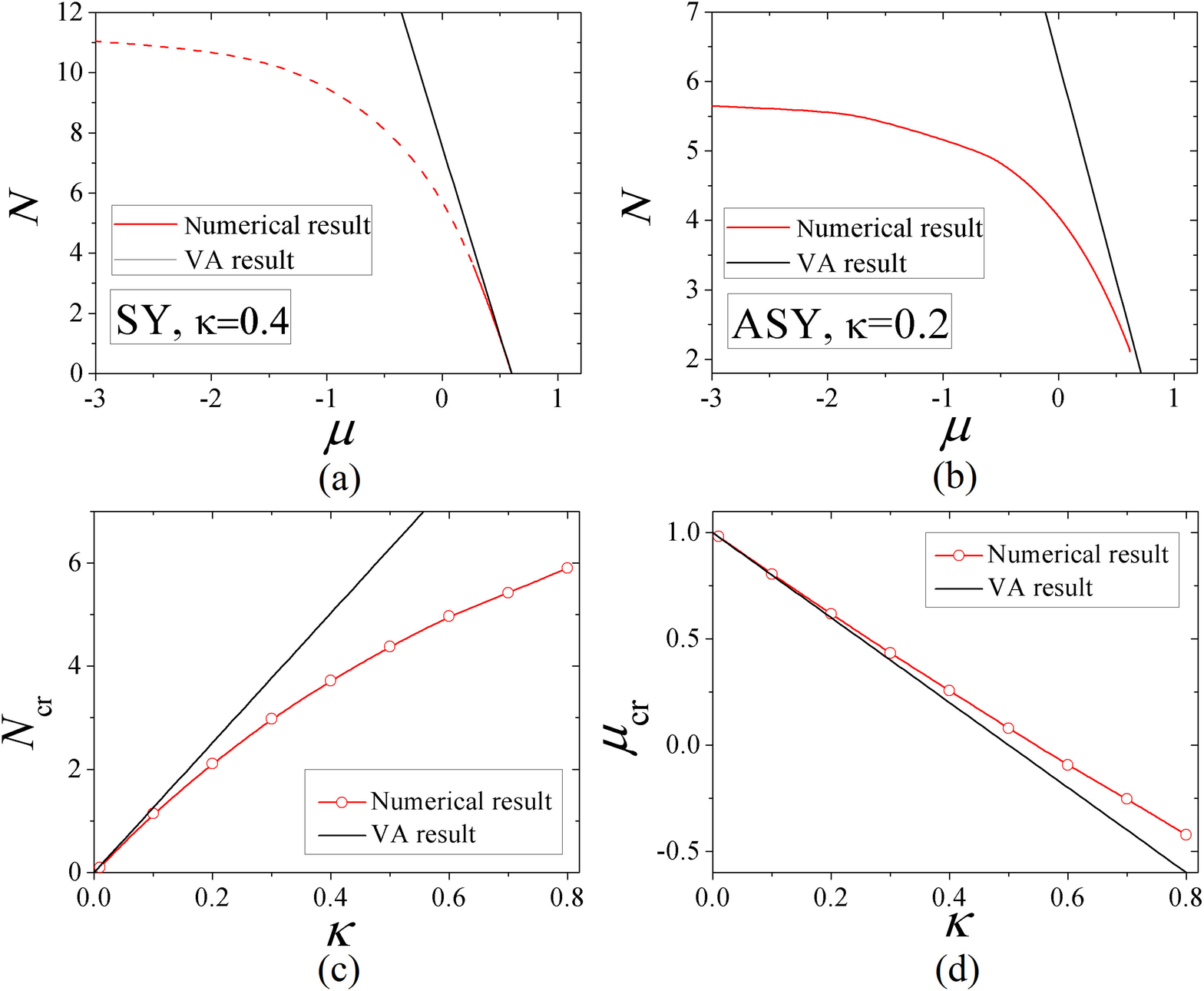}}
\caption{(Color online) Total norm $N$ versus chemical potential $\protect%
\mu $ for (a) symmetric (``SY") and (b) asymmetric (``ASY") ground states
(modes with $S=0$) in the 2D system, for indicated values of coupling
constant $\protect\kappa $. Red and black curves represent the numerical and
variational results, respectively, see Eqs. (\protect\ref{VAN2D}) and (%
\protect\ref{VAN2Dasymm}). The dashed segment of the numerically generated $%
N(\protect\mu )$ curve in (a) represents the branch destabilized by the SBB.
(c) Critical values of total norm $N_{\mathrm{cr}}^{(S=0)}$ at the
symmetry-breaking point of the GS in the 2D system, versus $\protect\kappa $.
(d) Critical values of the respective chemical potential, $\protect\mu _{%
\mathrm{cr}}^{(S=0)}$, versus $\protect\kappa $. Again, red and black curves
represent, severally, numerical and VA results, see Eq. (\ref{cr-2D}%
). The numerically found dependences displayed in (c) and (d) terminates at the point
determined by Eq. (\ref{kappa_max}).}
\label{FSNormchm}
\end{figure}

In a systematic form, the numerical results for the symmetric states are
presented, along with the respective VA prediction, in terms of the relation
between the total norm, $N$ [see Eq. (\ref{N2D})], and chemical potential, $%
\mu $, in Fig. \ref{FSNormchm}(a). The VA prediction stays close to the
numerical result when the nonlinearity is relatively weak [in particular,
the symmetric states displayed in Fig. \ref{syexample}(a) for $N=1$
correspond to a very small discrepancy in Fig. \ref{FSNormchm}(a)]. The
discrepancy increases as the nonlinearity grows stronger, causing, as said
above, the switch of the departure of the 2D wave function from the GS of
the HO towards the Townes soliton. In the limit of $\mu \rightarrow -\infty $%
, the total norm approaches value $N_{\max }^{(S=0)}\approx 11.70$, which is
twice the well-known norm of the Townes soliton, $N_{\mathrm{Townes}}\approx
5.85$, at which the critical collapse sets in \cite{Berge} (an approximate
variational prediction for it is $N_{\mathrm{Townes}}=2\pi $ \cite{Anderson}%
).

A typical $N(\mu )$ dependence for the 2D asymmetric GS is displayed in Fig. %
\ref{FSNormchm}(b), in which the numerically found and VA-predicted curves
commence at the respective SBB points. In the limit of $\mu \rightarrow
-\infty $, the total norm approaches the above-mentioned value $N_{\mathrm{%
Townes}}$, which asymptotically corresponds to the Townes soliton in
component $\Phi _{S=0}$, while the contribution from $\Psi _{S=0}$ vanishes,
as per Eq. (\ref{small}). Note that, in both cases shown in Figs. \ref%
{FSNormchm}(a) and (b), the $N(\mu )$ dependences satisfy the \textit{%
Vakhitov-Kolokolov} criterion, $dN/d\mu <0$, which is the well-known
necessary stability condition for modes supported by the self-focusing
nonlinearity \cite{VK,Berge}.

\begin{figure}[tbp]
\centering{\label{fig3a} \includegraphics[scale=0.18]{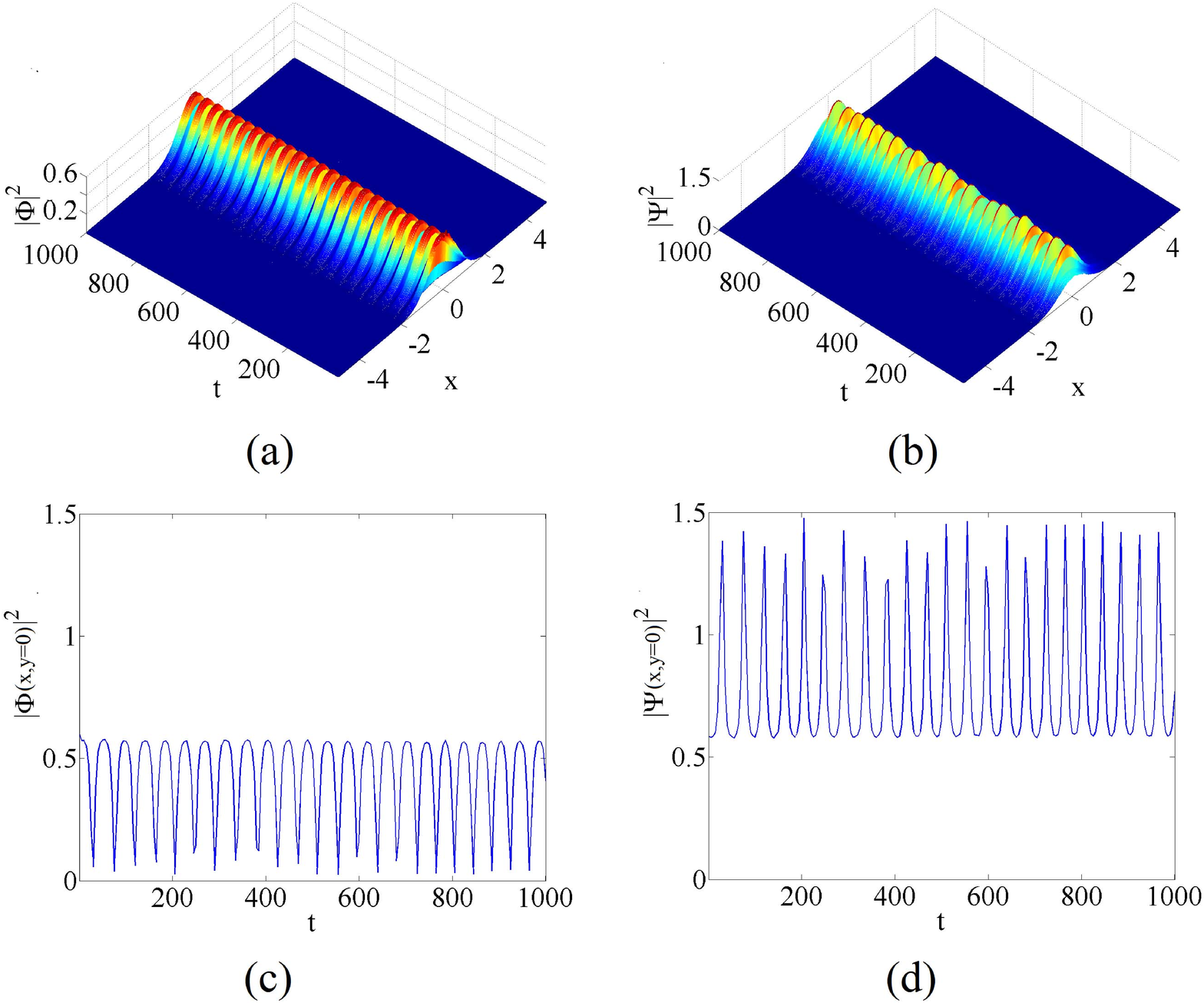}}
\caption{(Color online) (a,b) The numerically simulated evolution of an
unstable 2D symmetric fundamental ($S=0$) state (shown is cross-section $y=0$%
) with $(\protect\kappa ,N)=(0.2,3)$, which demonstrates the onset of the
spontaneous symmetry breaking with concomitant oscillations. (c,d) The
respective evolution of densities of the two components at the center.}
\label{syFsInOsci}
\end{figure}

Proceeding to detailed results for the stability of the symmetric and
asymmetric 2D fundamental modes (alias GSs), we note that, as it might be
expected, the asymmetric one is always stable when it exists (as the GS must
be). The symmetric state loses its stability beyond the SBB point, i.e., at $%
N>N_{\mathrm{cr}}^{(S=0)}$, see Eq. (\ref{cr-2D}). In direct simulations, it
spontaneously transforms into an asymmetric state with residual
oscillations, as shown in Fig. \ref{syFsInOsci}. Panels (c,d) in the figure
demonstrate that an amplitude minimum in one oscillating component is
attained simultaneously with the maximum in the other. This instability
occurs in the region of $N_{\mathrm{cr}}^{(S=0)}<N<N_{\mathrm{Townes}}$. At $%
N>N_{\mathrm{Townes}}$, the 2D\ system suffers the onset of the collapse, as
the spontaneous symmetry breaking makes the norm in one component much
larger than in the other, allowing the larger norm to reach the collapse
threshold for the single component, $N>N_{\mathrm{Townes}}$. As Fig. \ref%
{TempSyBrek} demonstrates, the symmetry breaking accelerates in the course
of the collapse development, driven by the growing nonlinearity strength in
the collapsing component, while the mate component suffers depletion, rather
than collapse.
\begin{figure}[tbp]
\centering{\label{fig4a} \includegraphics[scale=0.17]{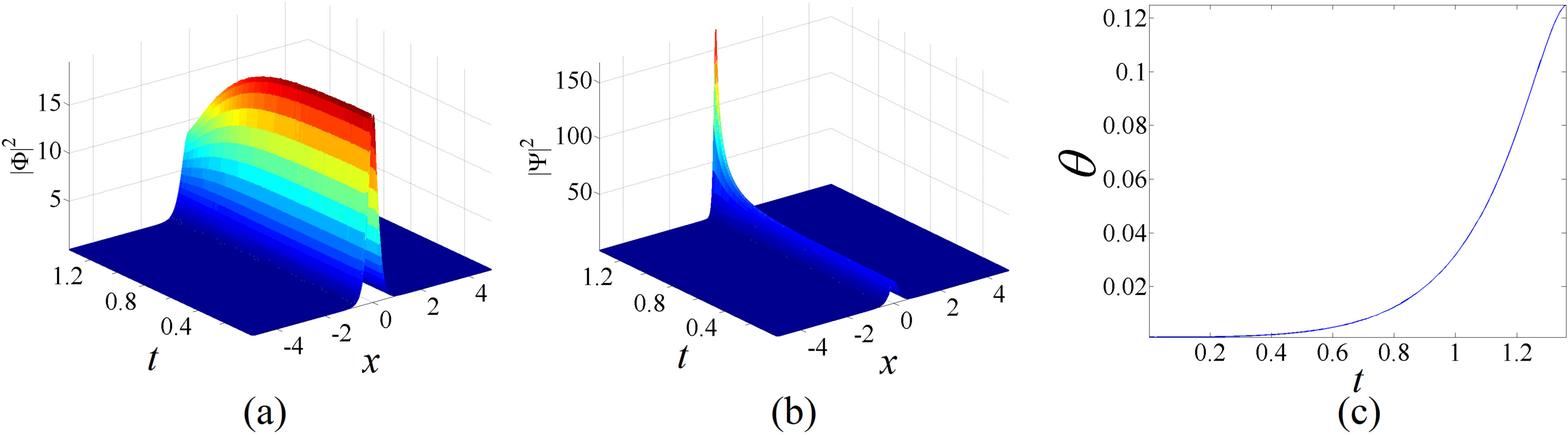}}
\caption{(Color online) (a,b) The evolution of an unstable symmetric
fundamental soliton (shown is cross-section $y=0$) with $(\protect\kappa %
,N)=(0.2,11.5)$, which demonstrates the acceleration of the symmetry
breaking in the course of the development of the collapse. In panel (c),
this is additionally illustrated by the dependence of asymmetry $\protect%
\theta $ between the two components [see Eq. (\protect\ref{theta})] on time
for the same solution.}
\label{TempSyBrek}
\end{figure}

\begin{figure}[tbp]
\centering{\label{fig5a} \includegraphics[scale=0.14]{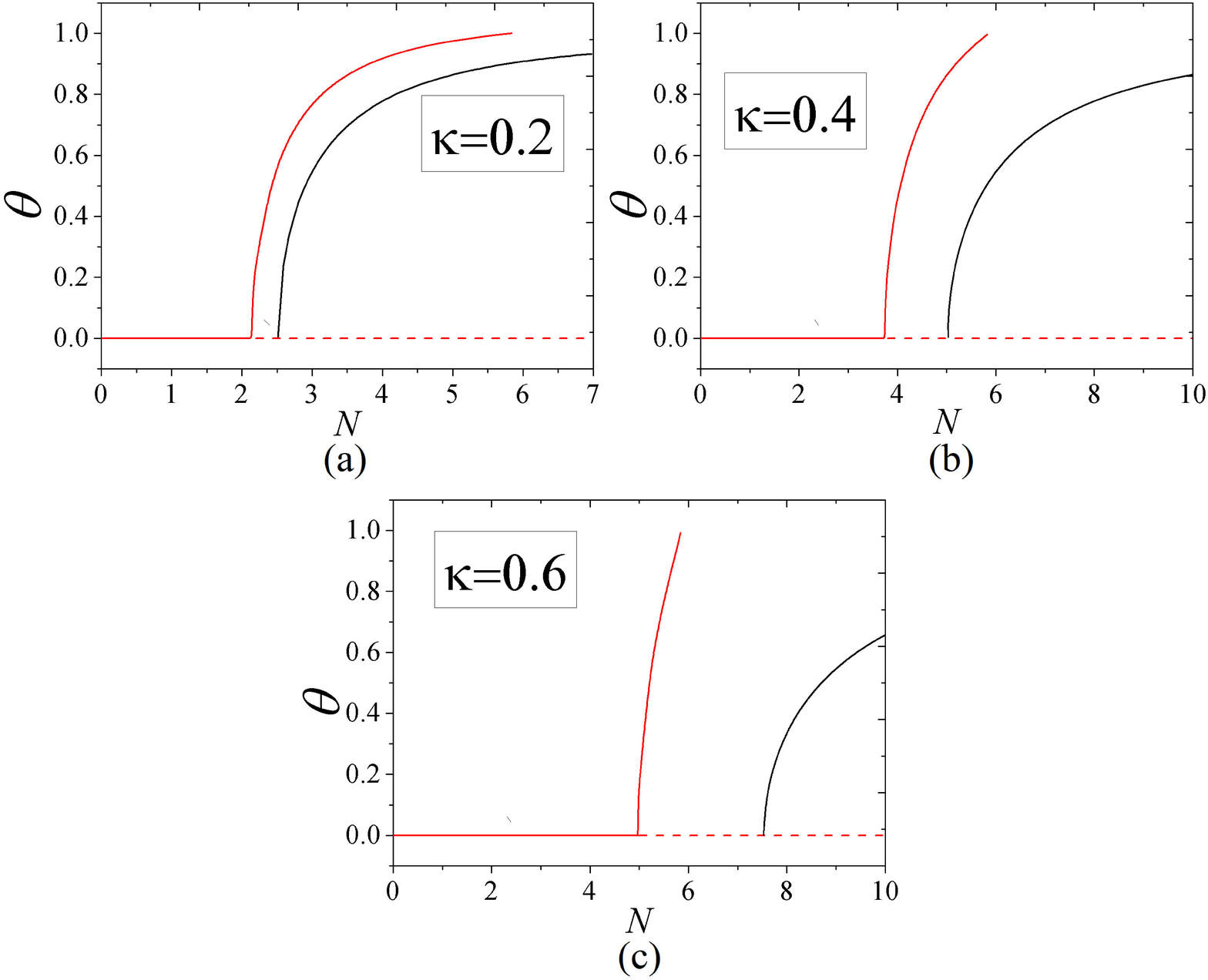}}
\caption{(Color online) Bifurcation diagrams, in the $(N,\protect\theta )$
plane, for the fundamental modes (ground states) with $S=0$, at different
values of the linear coupling constant: (a) $\protect\kappa =0.2$, (b) $%
\protect\kappa =0.4$, and (c) $\protect\kappa =0.6$. Here, black and
continuous/dashed red curves represent the variational results, and
numerically found stable/unstable solutions, respectively.}
\label{AsyNorm}
\end{figure}

The main objective of the analysis is the SBB in the two-component system.
Basic results for the bifurcation acting on the 2D fundamental modes are
presented in Fig. \ref{AsyNorm} in the form of relations between asymmetry $%
\theta $ [defined as per Eq. (\ref{theta})] and total norm $N$. In agreement
with the prediction of the VA, the corresponding SBB is of the supercritical
type. The numerically found $\theta (N)$ curves terminate at the
above-mentioned point of the onset of the critical collapse, $N=N_{\mathrm{%
Townes}}$.

With the increase of the coupling constant, $\kappa $, larger values of $N$
are required to for the onset of the symmetry breaking, therefore the
accuracy of the VA prediction for the SBB\ deteriorates at large $\kappa $.
It is relevant to stress that the SBB\ takes place in a finite interval of
the values of the coupling constant, $0<\kappa \leq \kappa _{\max }$, with $%
\kappa _{\max }$ determined by the condition that the symmetry breaking
occurs at $N=N_{\mathrm{Townes}}$, i.e., by equation
\begin{equation}
N_{\mathrm{cr}}^{(S=0)}\left( \kappa _{\max }\right) =N_{\mathrm{Townes}}
\label{max}
\end{equation}%
(at $\kappa >\kappa _{\max }$, the critical collapse occurs prior to the
expected onset of the SBB). The respective numerical result is
\begin{equation}
\kappa _{\max }^{(S=0)}\approx 0.8.  \label{kappa_max}
\end{equation}
It determines the termination points of the numerically found dependences $N_{{\rm cr}}(\kappa)$
and $\mu_{{\rm cr}}(\kappa)$ in Figs. \ref{FSNormchm}(c,d).

\subsubsection{Symmetric and asymmetric vortices with $S=1$}

\begin{figure}[tbp]
\centering{\label{fig6a} \includegraphics[scale=0.38]{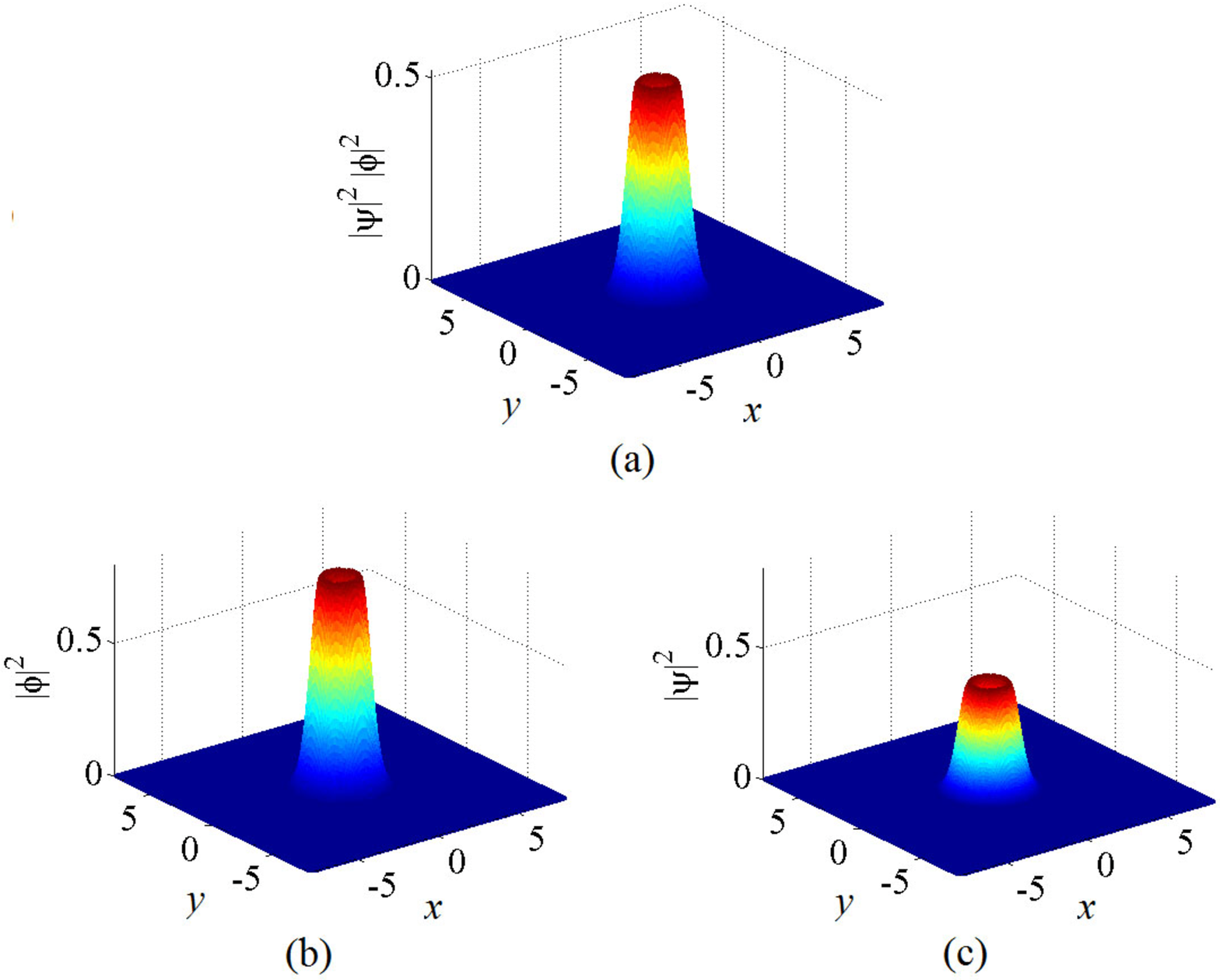}}
\caption{(Color online) (a) A stable symmetric vortex with $S=1$ and $(%
\protect\kappa ,N)=(0.4,8)$. (b,c) Two components of a stable asymmetric
vortex with $(\protect\kappa ,N)=(0.4,8.8)$. }
\label{syvotex1}
\end{figure}

\begin{figure}[tbp]
\centering{\label{fig7a} \includegraphics[scale=0.16]{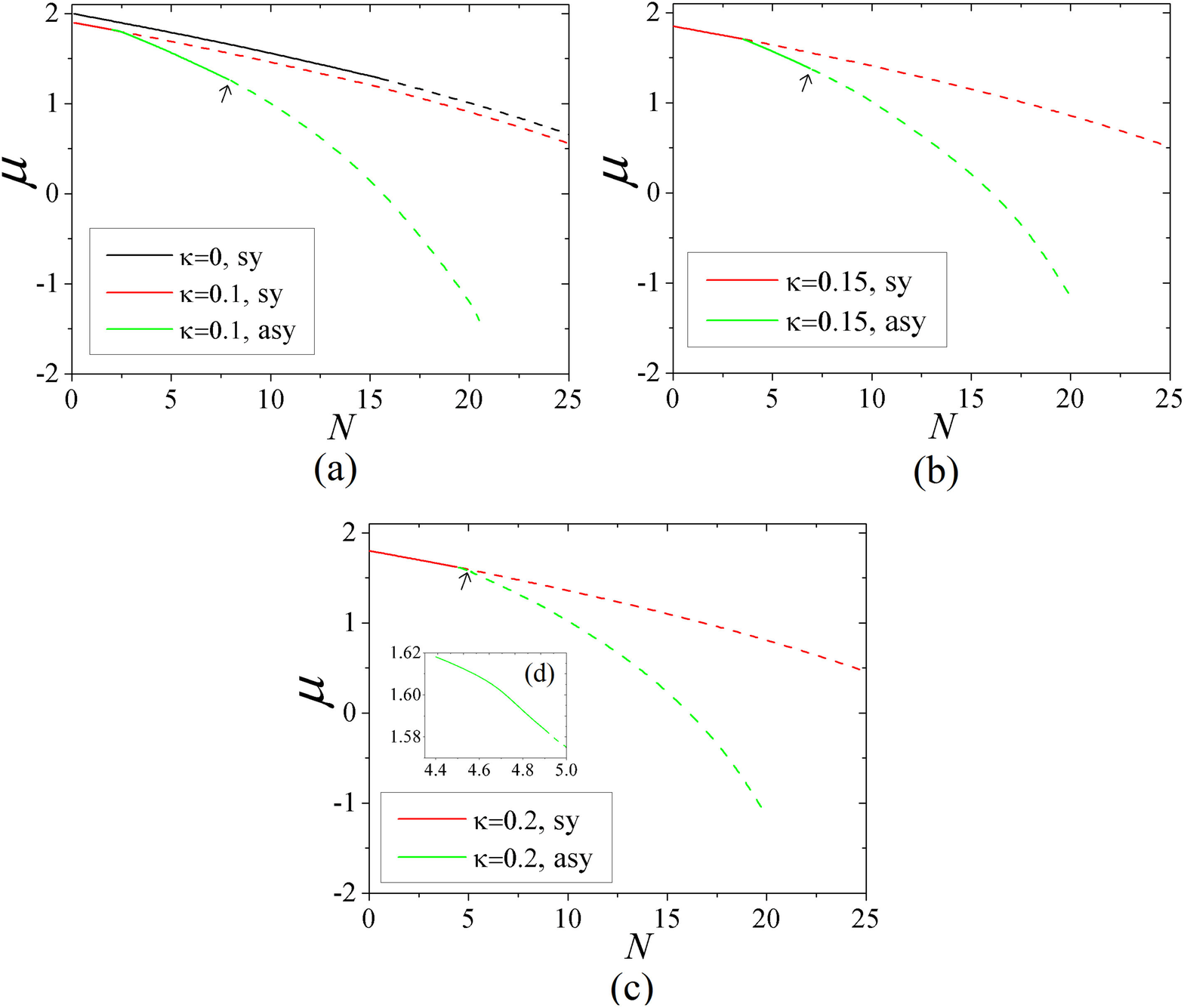}}
\caption{(Color online) Total norm $N$ versus chemical potential $\protect%
\mu $ for symmetric and asymmetric vortices with $S=1$, at indicated values
of the coupling constant, (a) $\protect\kappa=0.1$, (b) $\protect\kappa %
=0.15 $, (c) $\protect\kappa =0.2$. The decoupled system with $\protect\kappa%
=0$ is included in (a) too, for the completeness' sake. Inset (d) is a zoom
of a small stability region which exists in (c). Solid and dashed lines
designate stable and unstable branches, respectively. Arrows indicate points
where the asymmetric vortices loose their stability. They are $N_{\max
}^{(S=1)}$ for the asymmetric ones, see Eq. (\protect\ref{Nmax}).}
\label{SyAsyvMuN}
\end{figure}

Like GS, solutions of symmetric and asymmetric vortices are also produced by
means of imaginary-time-integration method. Typical examples of shapes of
stable symmetric vortices are shown in Figs. \ref{syvotex1}(a) and (b,c),
respectively. Relations between total norm and chemical potential for
families of vortices are displayed in Fig. \ref{SyAsyvMuN}, which shows that
the stability regions of the symmetric and asymmetric vortices expand and
shrink, respectively, with the increase of coupling constant $\kappa $. To
explain these findings, we note, first, that in the decoupled limit, $\kappa
=0$, the stability region for the vortices was known previously \cite{Dum}:
\begin{equation}
0\leq N\leq N_{\mathrm{cr}}^{(S=1)}\left( \kappa =0\right) \approx 15.6,
\label{decoupled}
\end{equation}%
which coincides with the stability boundary in Fig. \ref{SyAsyvMuN}(a) for $%
\kappa =0$ (it is multiplied by $2$ in comparison with Ref. \cite{Dum},
where the boundary was given for the single component).

\begin{figure}[tbp]
\centering{\label{fig8a} \includegraphics[scale=0.27]{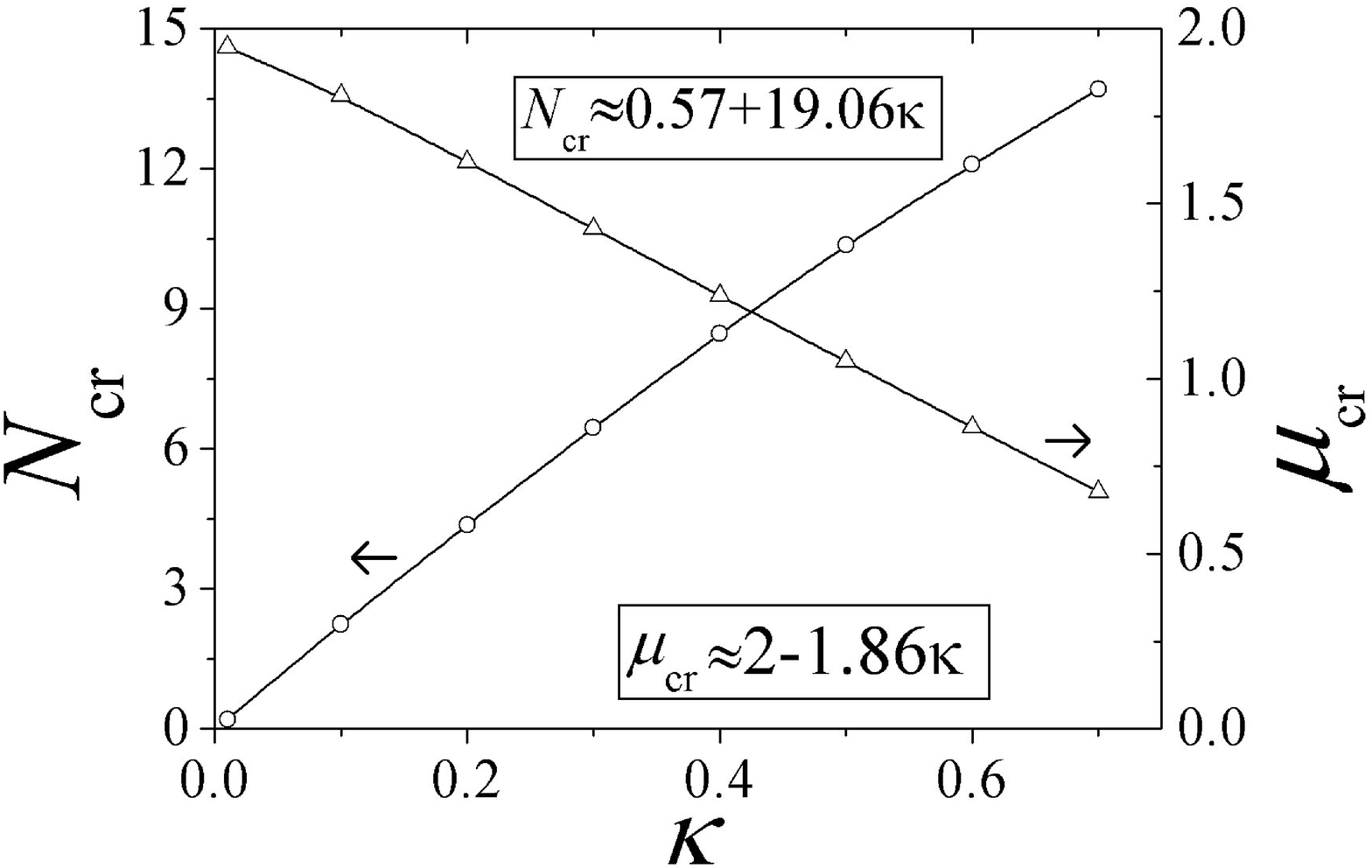}}
\caption{(Color online) The critical value of the total norm at the
symmetry-breaking point, $N_{\mathrm{cr}}^{(S=1)}$, versus the coupling
constant, $\protect\kappa $, for vortices with $S=1$, corresponds to the
black line with circles. The chemical potential at the bifurcation point, $%
\protect\mu _{\mathrm{cr}}^{(S=1)}$, versus the coupling constant $\protect%
\kappa $, is shown by the black line with triangles. As indicated in the
figure, these dependences can be fitted by linear relations (\protect\ref%
{emp}). They terminate at the point given by Eq. (\ref{kappa_max_S=1}).}
\label{Ncrvorx}
\end{figure}

Next, the symmetric vortices in the coupled system feature the SBB at the
respective critical value of the norm, $N_{\mathrm{cr}}^{(S=1)}(\kappa )$
[the $N(\mu )$ curves for the asymmetric vortices in Fig. \ref{SyAsyvMuN}(b)
for each $\kappa $ originate precisely at $N=N_{\mathrm{cr}}^{(S=1)}(\kappa
) $]. The numerically found dependence $N_{\mathrm{cr}}^{(S=1)}(\kappa )$ is
displayed in Fig. \ref{Ncrvorx}, along with the respective dependence on $%
\kappa $ of the chemical potential at the SBB point. They can be empirically
fitted by linear relations,
\begin{equation}
N_{\mathrm{cr}}^{(S=1)}(\kappa )=0.57+19.06\kappa ,~\mu _{\mathrm{cr}%
}^{(S=1)}(\kappa )=2-1.86\kappa .  \label{emp}
\end{equation}%
In fact, such linear dependences can be derived from the VA for the vortices
[cf. similar relations (\ref{cr-2D}) predicted by the VA for the modes with $%
S=0$]. We do not present this extension of the VA here in detail, as it is
rather cumbersome.

With the further increase of $\kappa $, the value $N_{\mathrm{cr}%
}^{(S=1)}(\kappa )$ attains the above-mentioned stability limit for the
decoupled system, given by Eq. (\ref{decoupled}). As follows from Eq. (\ref%
{emp}), this happens at
\begin{equation}
\kappa =\kappa _{\max }^{(S=1)}\allowbreak \approx 0.81,
\label{kappa_max_S=1}
\end{equation}
which is, interestingly, virtually the same as its counterpart for
the modes with $S=0$, given by Eq. (\ref{kappa_max}). Accordingly,
dependences $N_{\mathrm{cr}}^{(S=1)}(\kappa )$ and $\mu _{\mathrm{cr}%
}^{(S=1)}(\kappa )$, displayed in Fig. \ref{Ncrvorx}, terminate close to $%
\kappa =\kappa _{\max }^{(S=1)}$. At $\kappa >\kappa _{\max }^{(S=1)}$, no
stable asymmetric vortices exist, as the symmetric ones become unstable
prior to the onset of the SBB. This fact explains the shrinkage of the
stability region for the asymmetric vortices with the increase of $\kappa $,
as observed in Fig. \ref{SyAsyvMuN} and \ref{SyAsyRegion}.

\begin{figure}[tbp]
\centering{\label{fig9a} \includegraphics[scale=0.15]{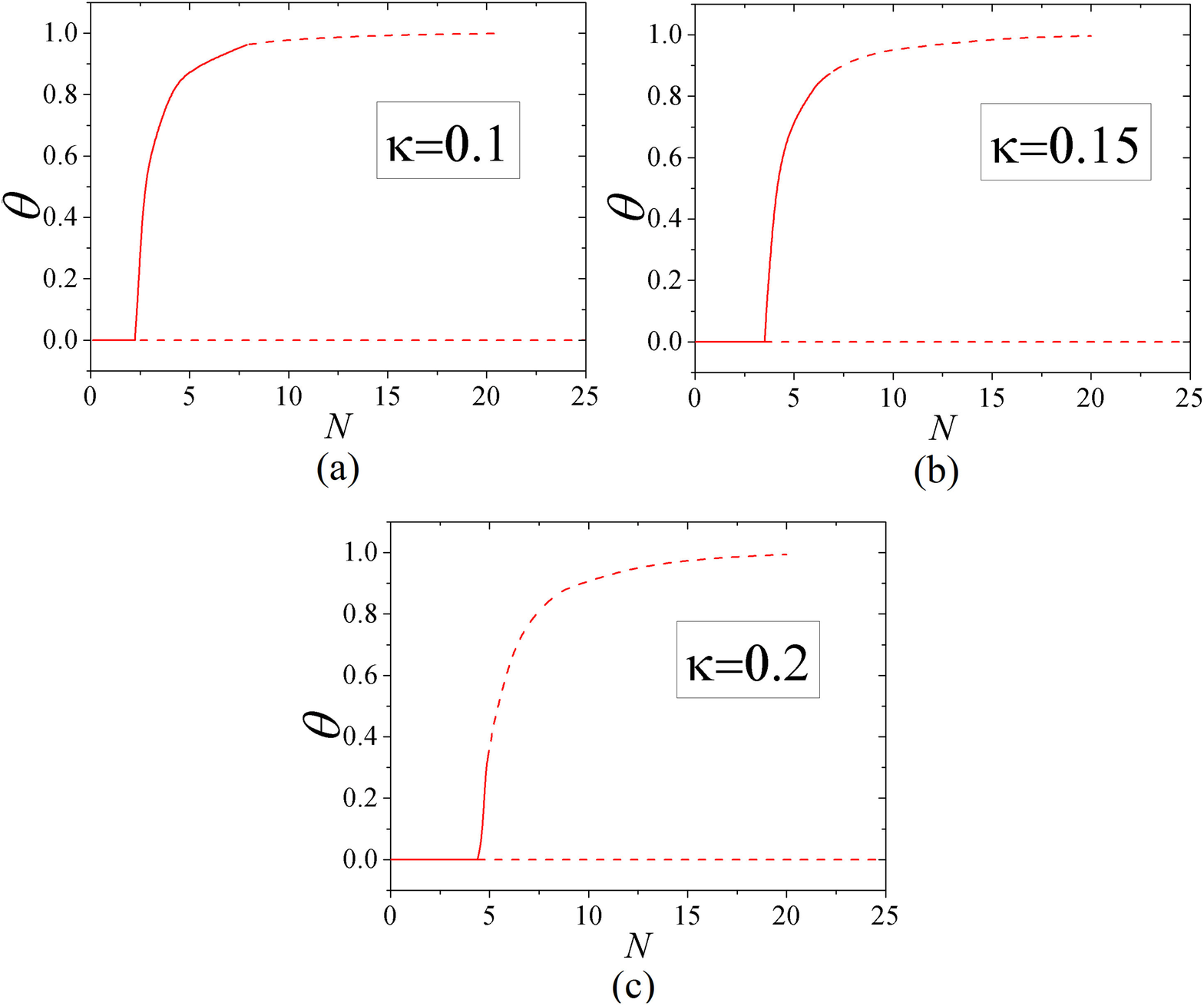}}
\caption{(Color online) Bifurcation diagrams, in the $(N,\protect\theta )$
plane, for the vortices with $S=1$ at different values of the linear
coupling constant: (a) $\protect\kappa =0.1$, (b) $\protect\kappa =0.15$,
and (c) $\protect\kappa =0.2$. Stable and unstable branches are shown by
solid and dashed lines, respectively. }
\label{bifurvorx}
\end{figure}

The SBB for vortices is illustrated by curves $\theta (N)$ which are
displayed, for different values of $\kappa $, in Fig. \ref{bifurvorx}, cf.
similar diagrams for the models with $S=0$ in Fig. \ref{AsyNorm}. These
diagrams clearly demonstrate that the the SBB for the vortices is also of
the supercritical type.

\begin{figure}[tbp]
\centering{\label{fig10a} \includegraphics[scale=0.19]{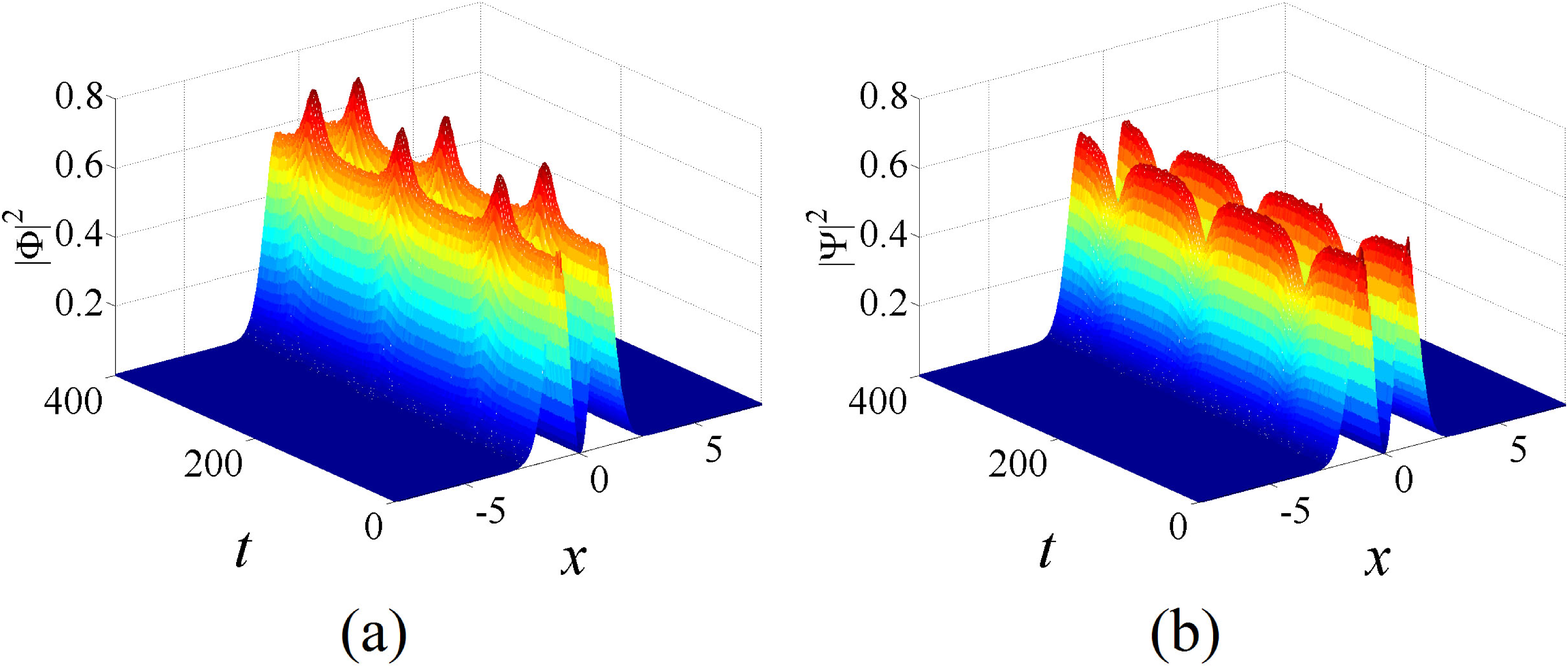}}
\caption{(Color online) The evolution of an unstable 2D symmetric ($S=1$)
vortex (shown is cross-section $y=0$) with $(\protect\kappa ,N)=(0.4, 8.8)$,
which demonstrates the onset of the spontaneously symmetry breaking with
residual oscillations. }
\label{VortSyInstb}
\end{figure}

Symmetric vortices destabilized by the SBB spontaneously transform into
asymmetric ones with residual oscillations(see Fig. \ref{VortSyInstb}),
similar to the spontaneous transition from unstable symmetric states with $%
S=0$ to oscillating asymmetric modes, as shown above in Fig. \ref{syFsInOsci}%
. Stable asymmetric vortices exist in the interval of $N_{\mathrm{cr}%
}^{(S=1)}<N<N_{\max }^{(S=1)}$, where $N_{\max }^{(S=1)}$ is the largest
norm up to which asymmetric vortices remain stable, as shown in Fig. \ref%
{SyAsyvMuN}:
\begin{equation}
N_{\max }^{(S=1)}(\kappa =0.1)\approx 7.9,~N_{\max }^{(S=1)}(\kappa
=0.15)\approx 6.9,~N_{\max }^{(S=1)}(\kappa =0.2)\approx 4.9.  \label{Nmax}
\end{equation}

\begin{figure}[tbp]
\centering{\label{fig11a} \includegraphics[scale=0.33]{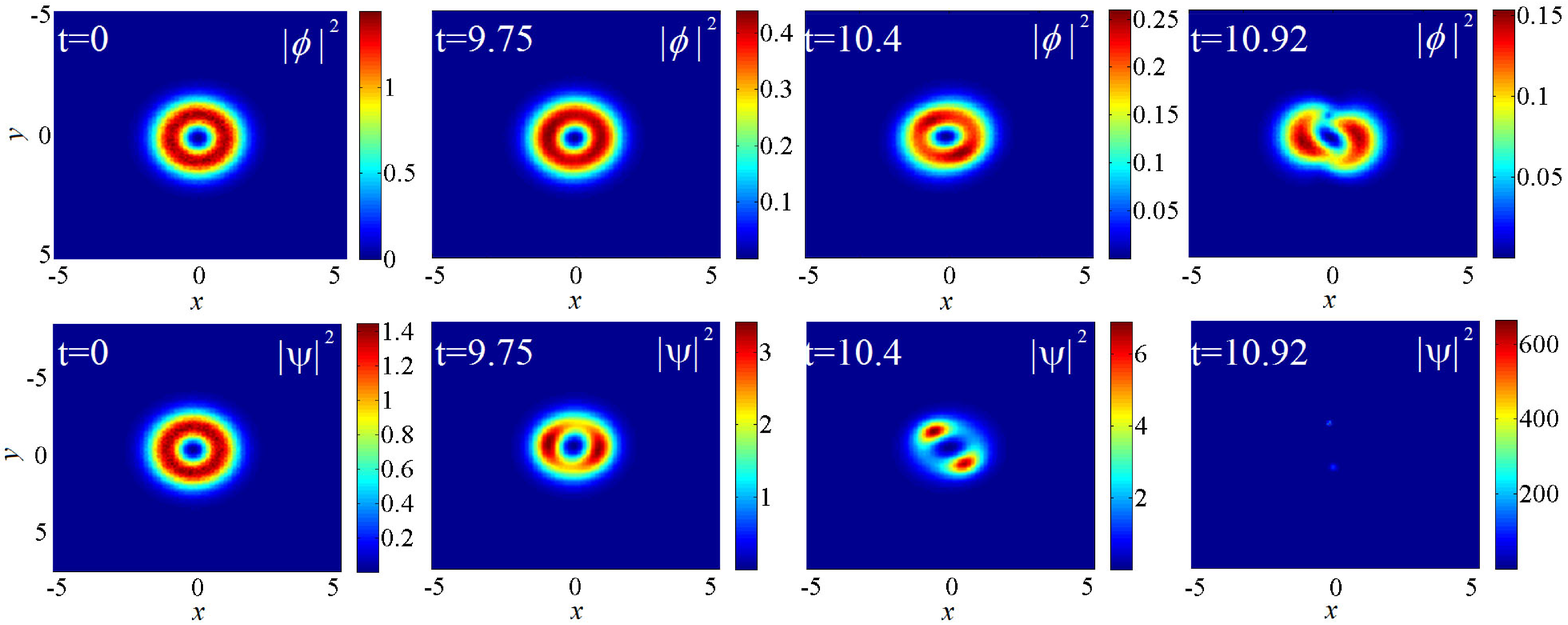}}
\caption{(Color online) Numerically simulated evolution of an unstable
symmetric vortex with $S=1$ and $(\protect\kappa ,N)=(0.8,18)$, initiated by
random noise (at the amplitude level of $5\%$) added to the input. This is a
typical example of the splitting instability.}
\label{SrcInstb}
\end{figure}

\begin{figure}[tbp]
\centering{\label{fig12a} \includegraphics[scale=0.31]{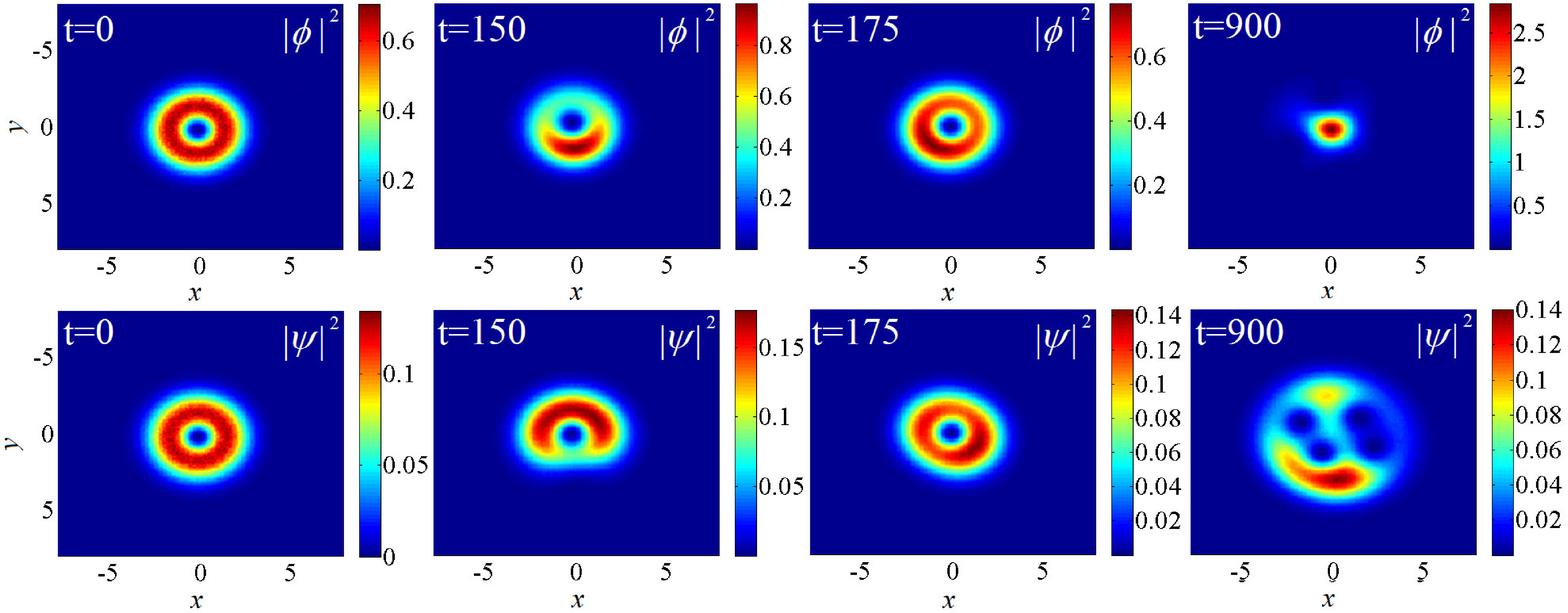}}
\caption{(Color online) Numerically simulated evolution of an unstable
asymmetric vortex with $S=1$ and $(\protect\kappa ,N)=(0.2,6)$, initiated by
random noise (at the amplitude level of $5\%$) added to the input. This
figure presents a typical example of the crescent instability (see the main
text).}
\label{AsyInstb}
\end{figure}

\begin{figure}[tbp]
\centering{\label{fig13a} \includegraphics[scale=0.18]{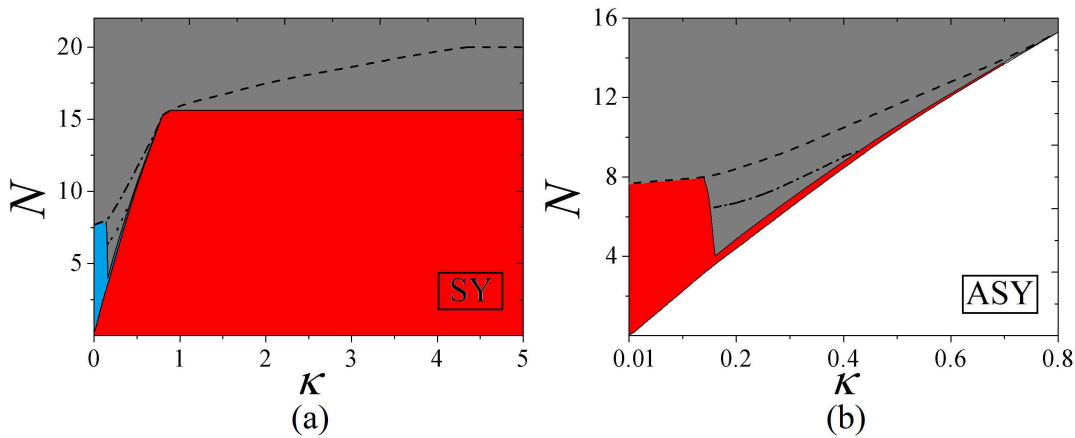}}
\caption{(Color online) The stability diagram for symmetric (a) and
asymmetric (b) vortices with $S=1$ in the plane of $(\protect\kappa,N)$. The
red and gray colors designate, respectively, stability and instability
areas. In panel (a), the small blue area represents instability with
residual oscillations, while the gray area below the dashed-dotted curve
designates the crescent instability similar to that shown in Fig. \protect
\ref{AsyInstb}. Above the dotted line, this instability leads to the
collapse of the component with the larger norm, but the collapse does not
happen in the small gray sub-area below the dotted line. The gray area above
the dashed-dotted and dashed lines represents the splitting instability scenario
displayed in Fig. \protect\ref{SrcInstb}, while robust recurrent
splitting-recombination cycles take place in the gray area below the dashed
curve. In panel (b), the gray area under the dashed-dotted line represents
the instability shown in Fig. \protect\ref{AsyInstb}, while in the area
between the dashed and dashed-dotted lines instability similar to that in
Fig. \protect\ref{AsyInstb} occurs, finally leading to the collapse.
Finally, the gray area above the dashed curve represents instability similar
to that displayed in Fig. \protect\ref{SrcInstb}.}
\label{SyAsyRegion}
\end{figure}

At $N>N_{\max }^{(S=1)}$, two generic instability scenarios are possible,
starting from the symmetric input. One (\textit{splitting})is shown in Fig. %
\ref{SrcInstb}, which demonstrates that the evolution of symmetric vortices
combines the spontaneous symmetry breaking between the $\phi $ and $\psi $
components and splitting of the vortex ring in two fragments, followed by
the collapse of the fragments in the component carrying a larger amplitude
(it is $\psi $, in Fig. \ref{SrcInstb}; the asymmetric collapse may be
compared to that displayed in Fig. \ref{TempSyBrek}\ for the modes with $S=0$%
). The second scenario (\textit{crescent instability})is illustrated by Fig. %
\ref{AsyInstb}: the asymmetric vortex rings spontaneously transforms into a
crescent, which is followed by recovery of the ring. After several cycles of
such transformations, the crescent's component with a larger amplitude tends
to evolve into an approximately fundamental state, while the component with
a smaller norm develops a chaotic pattern. The former component will
eventually suffer collapse, if the norm of the emergent fundamental state
exceeds the value corresponding to the collapse onset. A similar scenario of
the instability development of vortices in single-component self-attractive
BEC was reported in Ref. \cite{HSaito}.
%The evolution of unstable asymmetric vortices at $%
%N>N_{\max }^{(S=1)}$ also have two behaviours. One is similar to what is
%observed in Fig. \ref{SrcInstb} at the stage following the onset of the
%symmetry breaking: splitting into fragments, which subsequently suffer the
%collapse, asymmetrically with respect to the two components. The second one
%is shown in Fig. \ref{AsyInstb}, which is similar to its symmetric
%counterpart with same parameters $N$ and $\kappa $. Stability and
%instability diagrams for both symmetric and asymmetric vortices are
%demonstrated in Fig.\ref{SrcInstb}.

In the single-component model, simulations of the evolution of the vortex
with $S=1$ reveal a robust dynamical regime intermediate between the
stability and the splitting followed by the collapse: in the interval of the
norm which, in the present notation, is%
\begin{equation}
15.56<N<20,  \label{cycles}
\end{equation}%
the vortex ring recurrently splits into two fragments that recombine back
into the ring \cite{Dum,HSaito}. In the present system, systematic
simulations demonstrate that such a stable regime does not occur at values
of the coupling constant $0<\kappa \leq 0.81$, as the symmetry breaking
destabilizes the fragments after the first splitting (approximately in the
same fashion as shown in Fig. \ref{SrcInstb}). At $\kappa \geq 4.35$, the
linear coupling is so strong that the dynamics of the two-component system
is identical to that of its single-component counterpart, hence stable
splitting-recombination cycles, symmetric with respect to the $\phi $ and $%
\psi $ components, take place in interval (\ref{cycles}). In the case of $%
0.81<\kappa <4.35$, the splitting-recombination regime is stable below the
boundary in the $\left( \kappa ,N\right) $ plane, which is shown by the
dashed line in Fig. \ref{SyAsyRegion}(a).

\subsection{The one-dimensional system}

\begin{figure}[tbp]
\centering{\label{fig14a} \includegraphics[scale=0.17]{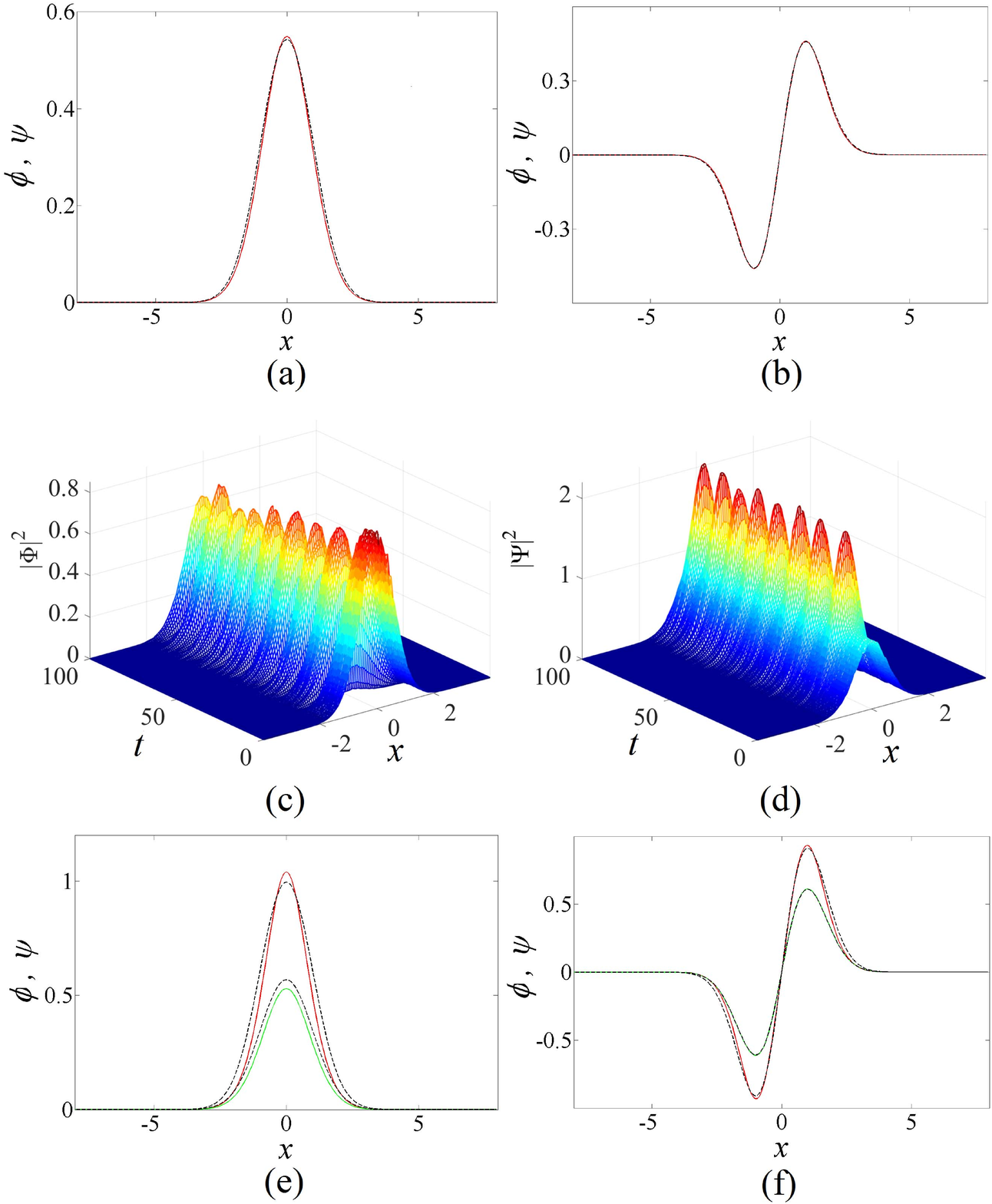}}
\caption{(Color online) (a,b) Typical examples of stable symmetric solutions
for the 1D ground-state and dipole modes, with $(\protect\kappa ,N)=(0.4,1)$%
. Red solid and black dashed curves display the numerical and variational
results, respectively. (c,d) The numerically simulated onset of the symmetry
breaking in an unstable symmetric ground-state mode, with $(\protect\kappa %
,N)=(0.4,2.5)$. Panels (e) and (f) display, severally, examples of stable
asymmetric ground-state solutions, with $\left( \protect\kappa ,N\right)
=(0.4, 2)$, and dipole modes, with $\left( \protect\kappa ,N\right) =(0.4,
2.7)$. Red and green solid curves represent two different components, as
produced by the numerical solution, while black dashed lines depict their
VA-predicted counterparts.}
\label{1Dexmpl}
\end{figure}

The VA for the 1D system, based on Eqs. (\ref{ansatz1D}) and (\ref{symm1D}),
(\ref{asymm1D}), predicts stable symmetric GS and dipole modes in a
virtually exact form, see typical examples in Fig. \ref{1Dexmpl}(a,b). Above
the SBB point, unstable symmetric GSs spontaneously transform into
asymmetric counterparts with residual oscillations, as shown in Fig. \ref%
{1Dexmpl}(c,d), cf. the similar transition in the 2D system, displayed above
in Fig. \ref{syFsInOsci}. The asymmetric GSs are completely stable in their
existence region.

\begin{figure}[tbp]
\centering{\label{fig15a} \includegraphics[scale=0.16]{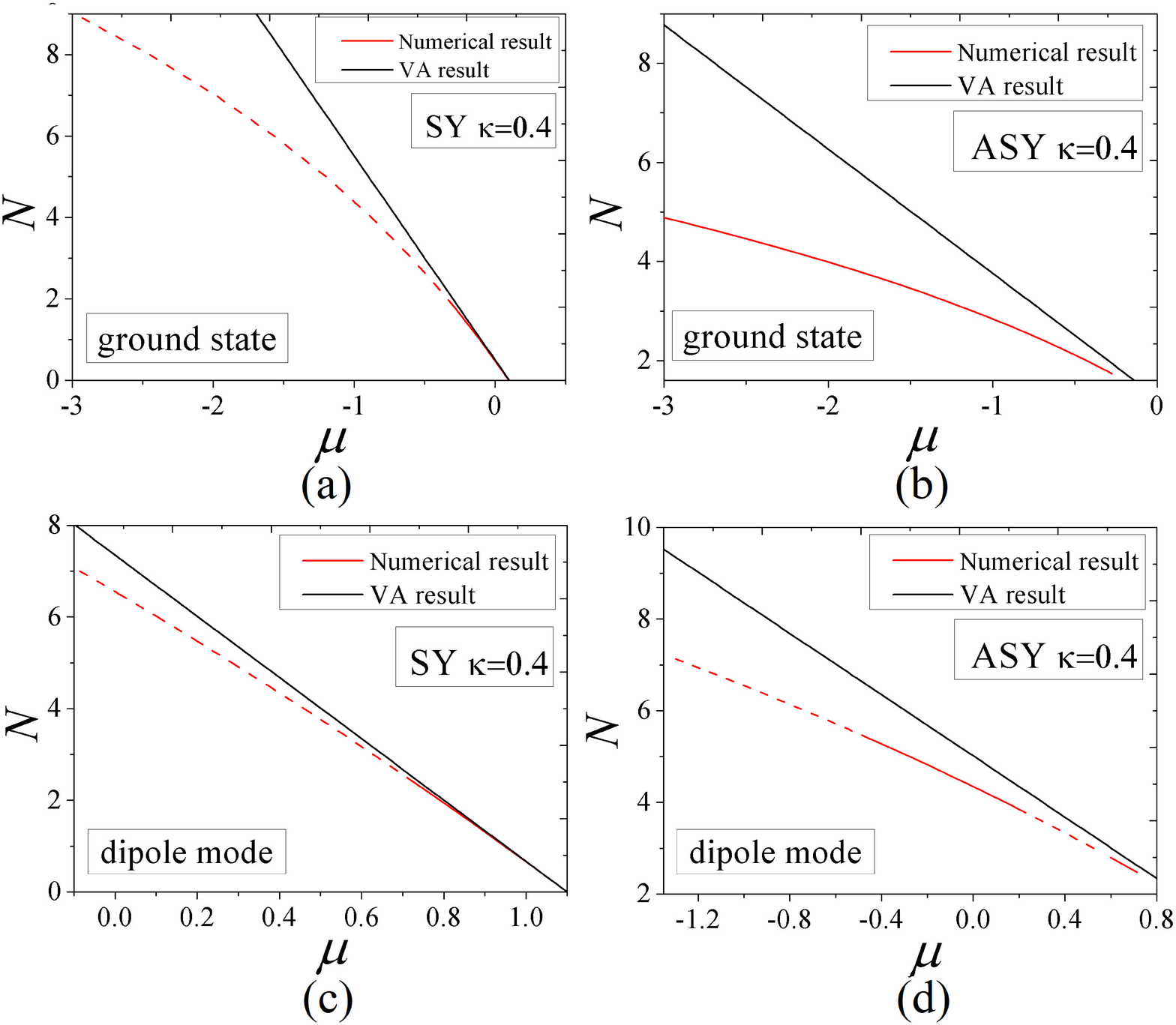}}
\caption{(Color online) Total norm $N$ versus chemical potential $\protect%
\mu $ for symmetric (\textquotedblleft SY") (a) and asymmetric
(\textquotedblleft ASY") (b) 1D ground state at $\protect\kappa =0.4$. Red
and black curves represent the numerical and variational results, the latter
given by Eqs. (\protect\ref{VAN1D}) \ and (\protect\ref{VAN1Dasymm}) for the
symmetric and asymmetric states, respectively. (c,d) The same for the 1D
dipole states, with the variational predictions given by Eqs. (\protect\ref%
{Nodd1Dtotal}) and (\protect\ref{Nodd1Dasymm}). Dashed are unstable segments
of the numerically generated branches. }
\label{1dNmu}
\end{figure}

\begin{figure}[tbp]
\centering{\label{fig16a} \includegraphics[scale=0.143]{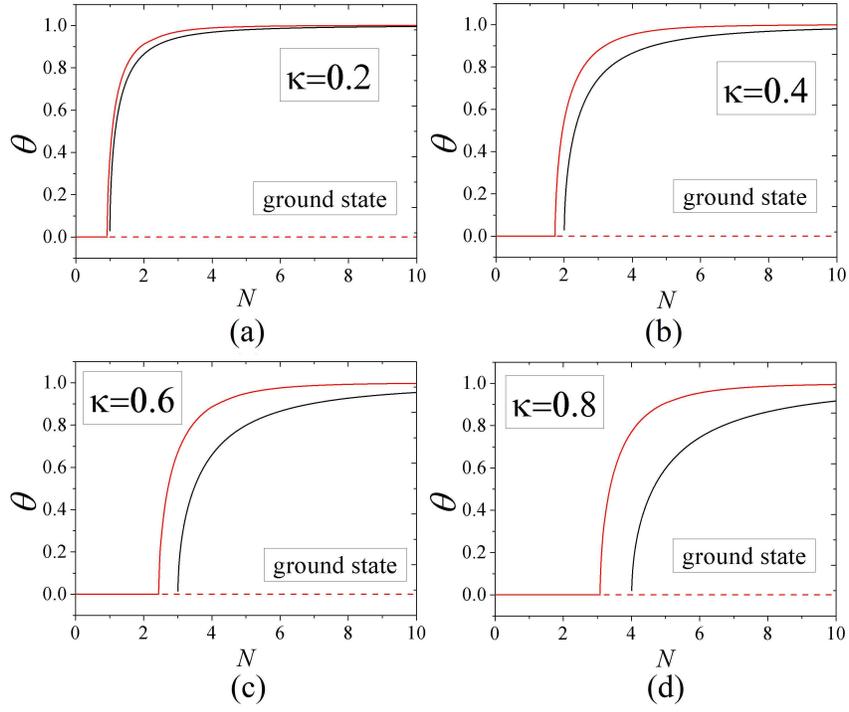}}
\caption{(Color online) Bifurcation diagrams, in the $(N,\protect\theta )$
plane, for the 1D ground-state modes, at different values of the
linear-coupling constant: (a) $\protect\kappa =0.2$, (b) $\protect\kappa %
=0.4 $, (c) $\protect\kappa =0.6$, and (d) $\protect\kappa =0.8$. The red
and black curves represent numerical findings and VA predictions,
respectively.}
\label{AsyN1dEven}
\end{figure}

\begin{figure}[tbp]
\centering{\label{fig17a} \includegraphics[scale=0.143]{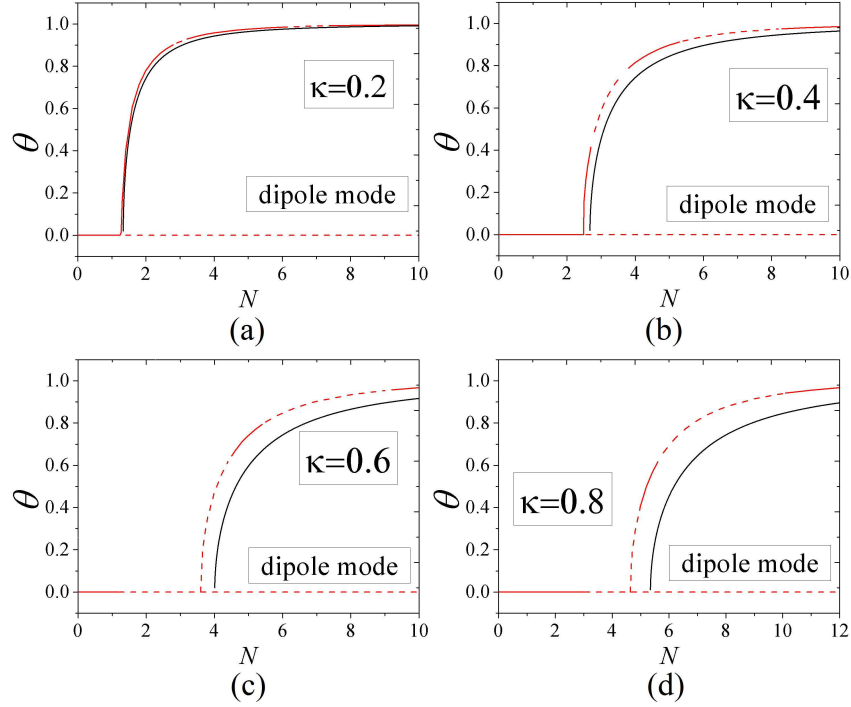}}
\caption{(Color online) The same as in Fig. \protect\ref{AsyN1dEven}, but
for the 1D dipole modes. }
\label{AsyN1dOdd}
\end{figure}

\begin{figure}[tbp]
\centering{\label{fig18a} \includegraphics[scale=0.16]{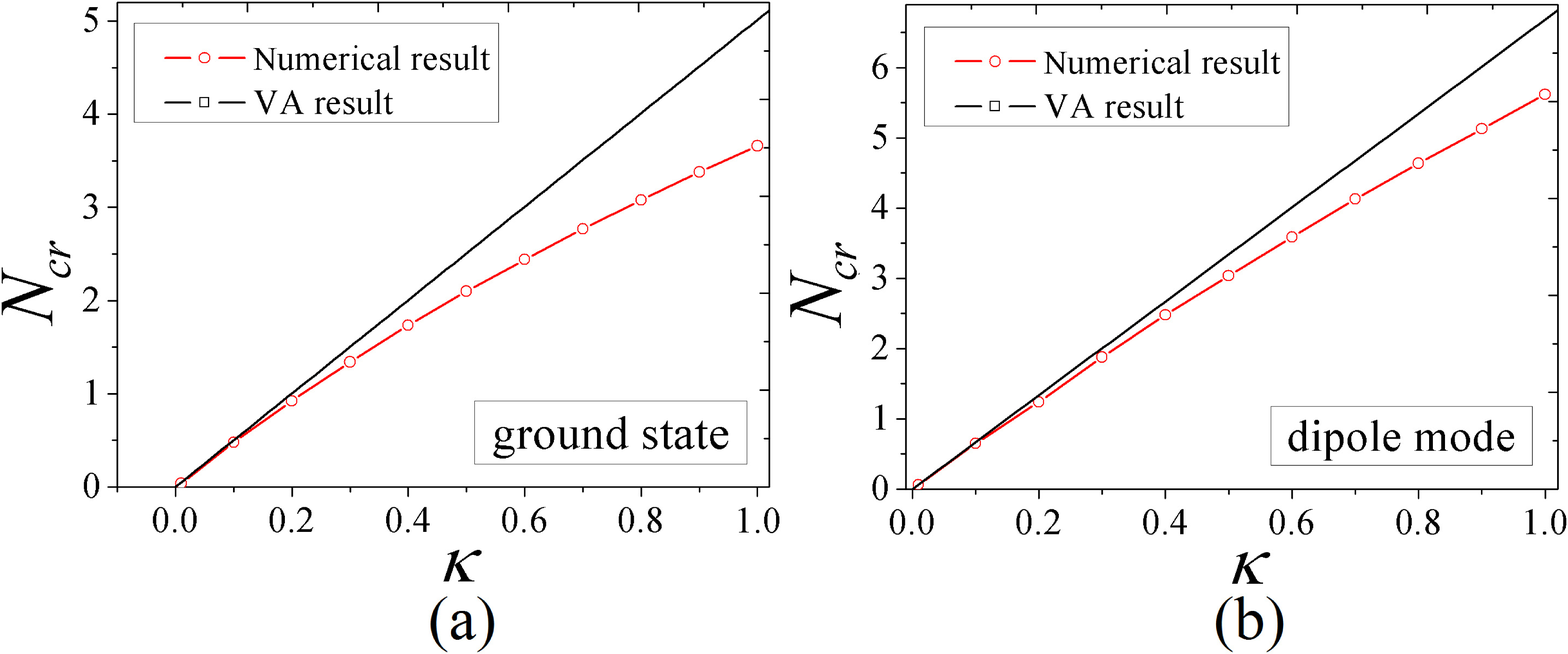}}
\caption{(Color online) (a) Critical values of total norm $N_{\mathrm{cr}}$,
at which the 1D ground state undergoes the symmetry-breaking transition,
versus coupling coefficient $\protect\kappa $. (b) The same for the 1D
dipole states. Again, red and black curves represent numerical and
variational results, respectively [see Eqs. (\protect\ref{cr-1D}) and (%
\protect\ref{Ncr-odd}), as concerns the VA predictions]. }
\label{NcrMcr1D}
\end{figure}

Systematic results for the 1D GS and dipole modes are reported by dint of $%
N(\mu )$ dependences in Fig. \ref{1dNmu}, and bifurcation diagrams in Figs. %
\ref{AsyN1dEven} and \ref{AsyN1dOdd}, respectively. In addition, critical
values of the norm at which the symmetry-breaking transition takes place for
both the GS and dipole modes are shown in Fig. \ref{NcrMcr1D}. The figures
also provide comparison of the VA predictions for these characteristics of
the solution families with the numerical findings. In agreement with the VA
result, the SBB in the 1D setting is of the supercritical type (on the
contrary to the weakly subcritical SBB for free-space 1D solitons in the
system of linearly coupled NLSEs \cite{bif1D}). Further, it is worthy to
note that the accuracy of the VA, based on \textit{ans\"{a}tze} (\ref%
{ansatz1D}) and (\ref{ansatz-odd}), is better for the dipole modes than for
the GS. This is explained by the fact that the intrinsic structure of the
dipoles makes them broader, hence their width is closer to that imposed by
the trapping potential, as implied by the \textit{ans\"{a}tze}, than to the
smaller soliton's width determined by the self-trapping. Another noteworthy
peculiarity revealed by Fig. \ref{NcrMcr1D} is that the critical norm is
somewhat higher for the dipoles than for the GS. This fact too is explained
by the effectively broader shape of the dipoles, which makes the
nonlinearity somewhat weaker for them, in comparison with the GS mode.

\begin{figure}[tbp]
\centering{\label{fig19a} \includegraphics[scale=0.2]{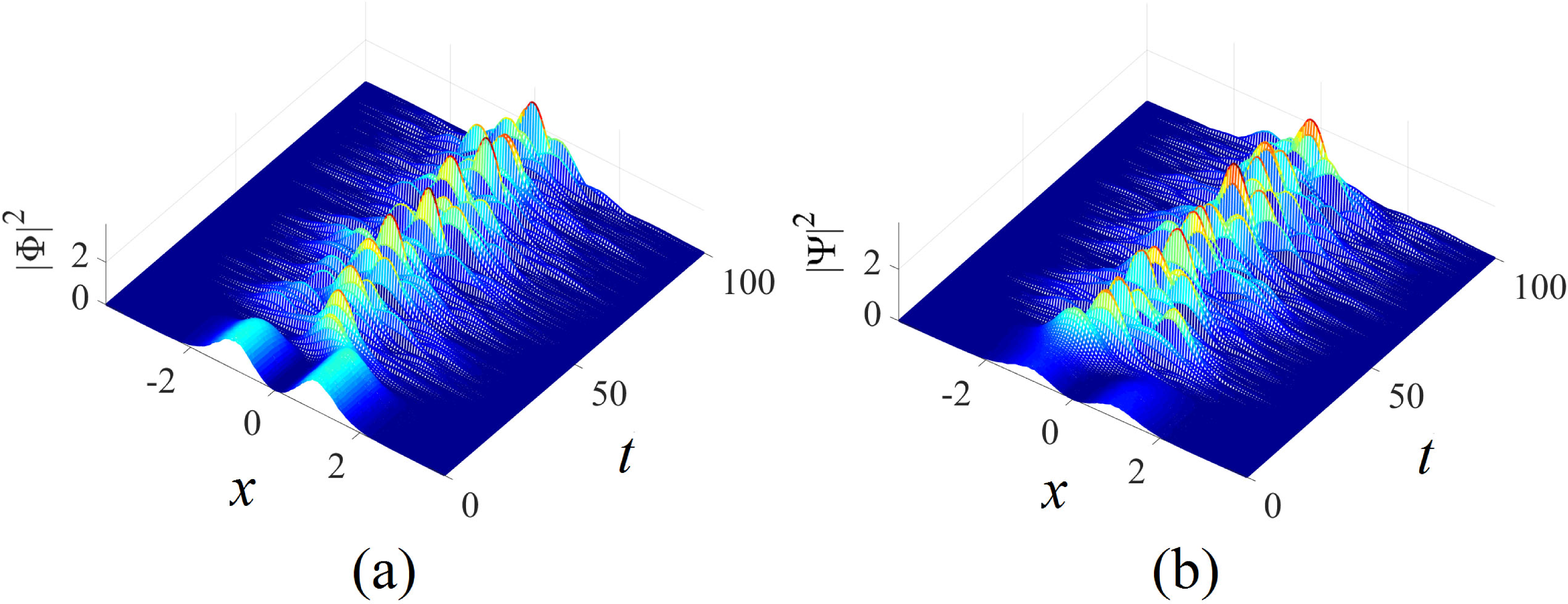}}
\caption{(Color online) The evolution of an unstable asymmetric dipole mode,
with $(\protect\kappa ,N)=(0.6,4)$, which transforms into a spatially
confined turbulent state.}
\label{EvenInstb}
\end{figure}

When 1D symmetric dipole modes are destabilized by the SBB, at the
respective critical points, $N=N_{\mathrm{cr}}^{\mathrm{(dip)}}$, they
spontaneously transform into asymmetric counterparts, featuring residual
oscillations (not shown here in detail), similar to the same transition for
the GS, cf. Fig. \ref{1Dexmpl}(c,d). A more interesting finding, specific to
the dipole mode, is that the branches of the asymmetric solutions, which are
completely stable in the case of the 1D GSs, feature instability segments,
with respective complex instability eigenvalues, corresponding to an
oscillatory instability. As seen in Figs. \ref{1dNmu}(d) and \ref{AsyN1dOdd}%
, these segments are relatively narrow for small $\kappa $, essentially
expanding at larger $\kappa $. In particular, at $\kappa =0.2$ and $0.4$,
the asymmetric dipole states remain stable, respectively, in intervals $N_{%
\mathrm{cr}}^{\mathrm{(dip)}}(\kappa =0.2)=1.24<N<2.9$ and $N_{\mathrm{cr}}^{%
\mathrm{(dip)}}(\kappa =0.4)=2.48<N<2.7$ [see Figs. \ref{AsyN1dOdd}(a,b)].
Furthermore, Figs. \ref{AsyN1dOdd} (c,d) demonstrate that, at $\kappa =0.6$
and $0.8$, this instability expands to a part of the originally stable
branches of the symmetric dipole modes below the SBB point. Direct
simulations show, in Fig. \ref{EvenInstb}, that this specific type of the
instability of the asymmetric and symmetric dipole modes transforms them
into robust chaotically oscillating states. This instability is not
essentially related to the two-component structure of the system, as the
emerging chaotic states seem effectively symmetric, with respect to the two
components.

\section{Conclusion}

The objective of this work is to study manifestations of the spontaneously
symmetry breaking of 2D fundamental modes and vortices, as well of 1D GS
(ground state) and dipole mode (the first excited state), in the
two-component model based on linearly coupled GPEs/NLSEs
(Gross-Pitaevskii/nonlinear Schr\"{o}dinger equations), with cubic
self-attraction and the HO (harmonic-potential) trap. The system can be
implemented in a two-layer BEC, and the 1D version also applies to dual-core
optical waveguides. Although effects of the spontaneously symmetry breaking
were studied, theoretically and experimentally, in many settings, these
quite fundamental realizations were not considered before.

Families of 2D fundamental and vortical states, with vorticities $S=0$ and $%
S=1$, have been constructed by means of the variational and numerical
methods, and their stability has been investigated by means of the
computation of eigenvalues for small perturbations, and further verified in
direct simulations. The respective SBB (symmetry-breaking bifurcation) is of
the supercritical type (i.e., it represents a phase transition of the second
kind) in all the cases. The asymmetric fundamental modes are completely
stable, while asymmetric vortices keep their stability in rather narrow
intervals of values of the norm, $N$, suffering the splitting into two
fragments and subsequent collapse of the fragments at larger $N$, or
featuring several cycles of transformations between the vortex ring and
crescent, and eventually transforming into a fundamental state in one
component, and a chaotic state in the other. The SBB occurs if the
linear-coupling constant, $\kappa $, falls below a certain value $\kappa
_{\max }\approx 0.81$, which is practically the same for the modes with $S=0$
and $S=1$. At $\kappa >\kappa _{\max }$, the strongly-coupled system
behaves, essentially, as the single-component one; in particular, vortices
demonstrate the regime of stable splitting-recombination cycles.

In 1D, families of GS and dipole solutions were found too in the variational
and numerical forms. They also demonstrate the SBB of the supercritical
type. The dipole modes are especially interesting, as, unlike the GSs and 2D
vortices, they do not exist in the free space, and they are better
approximated by the variational ansatz. A noteworthy feature of the
asymmetric and symmetric dipoles is the appearance, with the increase of $%
\kappa $, of additional regions of oscillatory instability, unrelated to the
spontaneous symmetry breaking. This instability transforms the dipoles into
confined turbulent states.

A subject for continuation of the present analysis may be the study of Rabi
oscillations between the two linearly components, in the 2D and 1D
geometries alike, cf. Ref. \cite{double-well-BEC}. It may also be
interesting to consider a generalization for a system with a spinor (binary)
wave function, which, in particular, may feature linear interconversion
between spinor components inside each layer. In the latter context, the
consideration of the spinor wave function subject to spin-orbit coupling
\cite{SOC} should be quite relevant. Lastly, it may be interesting too to
introduce a $\mathcal{PT}$-symmetric version of the system \cite{PT}, with
equal amounts of linear gain and loss added to the two coupled equations,
cf. Ref. \cite{Gena}.

\section*{Acknowledgments}

This work was supported, in part, by grant No. 2015616 from the joint
program in physics between NSF and Binational (US-Israel) Science
Foundation, grant No. 1286/17 from the Israel Science Foundation, by
project BIRD164754 of the University of Padova, and by NNSFC (China)
through Grant No. 11575063. Z.C. appreciates an excellence scholarship
provided by the Tel Aviv University.

\end{document}